\documentclass[preprint,journal]{vgtc}       % preprint (journal style)

\usepackage[dvipsnames]{xcolor}
\usepackage{amsmath,amsfonts,amssymb}
\usepackage[english]{babel}    % required for renaming appendix
\usepackage{chngcntr}          % required for renaming figure labels in appendix

\usepackage{comment}
\usepackage{gensymb}  % for \degree symbol
\usepackage{booktabs} % pretty tables
\usepackage{siunitx} % for setting precision in tables

\ieeedoi{10.1109/TVCG.2019.2934536}
\manuscriptnote{$\copyright$ \the\year~IEEE}

\addto{\captionsenglish}{}

\newcommand{\figskip}{\vspace{-3mm}}

\newcommand{\R}{\mathbb{R}}

\newcommand{\parsec}[1]{\vspace{1mm}\noindent\textbf{#1}\,}

\DeclareMathAlphabet{\mathsfit}{T1}{\sfdefault}{\mddefault}{\sldefault} % for italic sans-serif font in math mode

\ifpdf%                                % if we use pdflatex
  \pdfoutput=1\relax                   % create PDFs from pdfLaTeX
  \usepackage{graphicx}                % allow us to embed graphics files
  \DeclareGraphicsExtensions{.pdf,.png,.jpg,.jpeg} % for pdflatex we expect .pdf, .png, or .jpg files
\else%                                 % else we use pure latex
  \usepackage{graphicx}                % allow us to embed graphics files
  \DeclareGraphicsExtensions{.eps}     % for pure latex we expect eps files
\fi%

\graphicspath{{figures/}{pictures/}{images/}{./}} % where to search for the images

\usepackage{microtype}                 % use micro-typography (slightly more compact, better to read)
\PassOptionsToPackage{warn}{textcomp}  % to address font issues with \textrightarrow
\usepackage{textcomp}                  % use better special symbols
\usepackage{mathptmx}                  % use matching math font
\usepackage{times}                     % we use Times as the main font
\usepackage{cite}                      % needed to automatically sort the references
\usepackage{tabu}                      % only used for the table example
\usepackage{booktabs}                  % only used for the table example
\onlineid{0}

\vgtccategory{Research}
\vgtcpapertype{Evaluation}

\title{Estimating Color-Concept Associations from Image Statistics}

\author{Ragini Rathore, Zachary Leggon, Laurent Lessard, and Karen B. Schloss}

\authorfooter{
\item Ragini Rathore, Computer Sciences and Wisconsin Institute for Discovery (WID), University of Wisconsin--Madison,  Email: rrathore3@wisc.edu.
\item Zachary Leggon, Biology and WID, University of Wisconisn--Madison,  Email: zleggon@wisc.edu.
\item  Laurent Lessard, Electrical and Computer Engineering and WID, University of Wisconsin--Madison. Email: laurent.lessard@wisc.edu.
\item  Karen B. Schloss, Psychology and WID, University of Wisconsin--Madison. Email: kschloss@wisc.edu.
}

\shortauthortitle{Rathore \MakeLowercase{\textit{et~al.}}: Estimating Color-Concept Associations from Image Statistics}
\abstract{To interpret the meanings of colors in visualizations of categorical information, people must determine how distinct colors correspond to different concepts. This process is easier when assignments between colors and concepts in visualizations match people's expectations, making color palettes \textit{semantically interpretable}. Efforts have been underway to optimize color palette design for semantic interpretablity, but this requires having good estimates of human color-concept associations. Obtaining these data from humans is costly, which motivates the need for automated methods. We developed and evaluated a new method for automatically estimating color-concept associations in a way that strongly correlates with human ratings. Building on prior studies using Google Images, our approach operates directly on Google Image search results without the need for humans in the loop. Specifically, we evaluated several methods for extracting raw pixel content of the images in order to best estimate color-concept associations obtained from human ratings. The most effective method extracted colors using a combination of cylindrical sectors and color categories in color space. We demonstrate that our approach can accurately estimate average human color-concept associations for different fruits using only a small set of images. The approach also generalizes moderately well to more complicated recycling-related concepts of objects that can appear in any color.}


\CCScatlist{ % not used in journal version
 \CCScat{K.6.1}{Management of Computing and Information Systems}%
{Project and People Management}{Life Cycle};
 \CCScat{K.7.m}{The Computing Profession}{Miscellaneous}{Ethics}
}

\teaser{
  \centering
  \includegraphics[width=\linewidth]{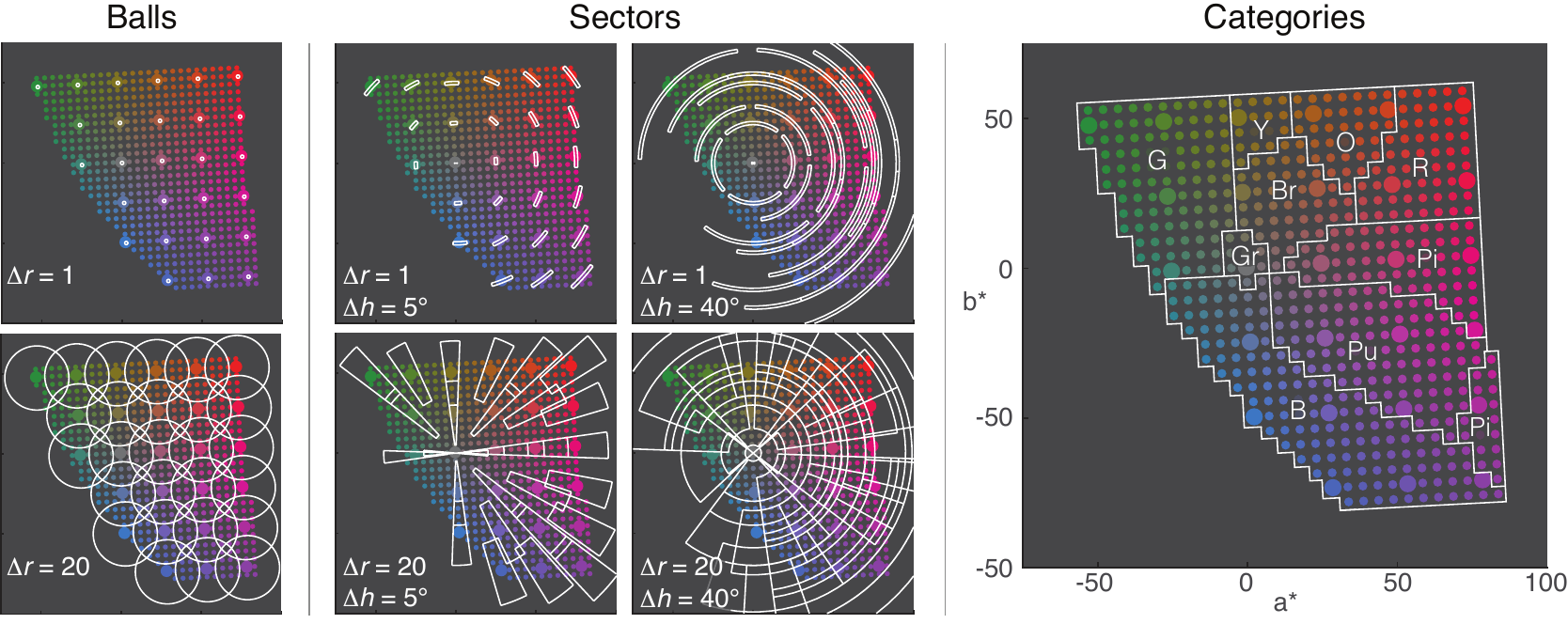}\figskip
  \caption{We constructed models that estimate human color-concept associations using color distributions extracted from images of relevant concepts. We compared methods for extracting color distributions by defining different kinds of color tolerance regions (white outlines) around each target color (regularly spaced large dots) in CIELAB space.  
  Subplots show a planar view of CIELAB space at L*\,=\,50, with color tolerance regions defined as balls (left column; radius $\Delta \mathsfit{r}$), cylindrical sectors (middle columns; radius $\Delta \mathsfit{r}$ and hue angle $\Delta \mathsfit{h}$), and category boundaries around each target color (right column; \textbf{R}ed, \textbf{O}range, \textbf{Y}ellow, \textbf{G}reen, \textbf{B}lue, \textbf{Pu}rple, \textbf{Pi}nk, \textbf{Br}own, \textbf{Gr}ay; white and black not shown).
  Each target color is counted as ``present'' in the image each time any color in its tolerance region is observed. This has a smoothing effect, which enables the inclusion of colors that are not present in the image but similar to colors that are. A model that includes two sector features and a category feature best approximated human color-concept associations for unseen concepts and images (see text for details).
}
	\label{fig:CirclesSectors}
}

\vgtcinsertpkg

\begin{document}


\firstsection{Introduction}

\maketitle
In visualizations of categorical information (e.g., graphs, maps, and diagrams), designers encode categories using visual properties (e.g.,  colors, sizes, shapes, and textures) \cite{bertin1983}. Color is especially useful for encoding categories for two main reasons. First, cognitive representations of color have strong categorical structure \cite{rosch1973, berlin1969, witzel2018}, which naturally maps to categories of data \cite{healey1996, brewer1994}. Second, people have rich semantic associations with colors called \textit{color-concept associations} (e.g., a particular red associated with strawberries, roses, and anger) \cite{osgood1957, humphrey1976, palmer2010}, which they use to interpret meanings of colors in visualizations \cite{lin2013, setlur2016, schloss2018, schloss2019, kinateder2019}. Indeed, it is easier to interpret visualizations if semantic encoding between colors and concepts (referred to as \textit{color-concept assignments)} match people's expectations derived from their color-concept associations \cite{lin2013, schloss2018, schloss2019}.  

Recent research has investigated how to optimize color palette design to produce color-concept assignments that are easy to interpret\cite{lin2013, setlur2016, schloss2018}. Methods typically involve  quantifying associations between each color and concept of interest, and then using those data to calculate optimal color-concept assignments for the visualization \cite{lin2013, schloss2018, bartram2017, setlur2016}. It may seem that the best approach would be to assign concepts to their most strongly associated colors, but that is not always the case. Sometimes, it is better to assign concepts to weakly associated colors to avoid confusions that can arise when multiple concepts are associated with similar colors (see Section \ref{sec:interpreting}) \cite{schloss2018}.  Thus, to leverage these optimization methods for visualization design, it is necessary to have an effective and efficient way to quantify human color-concept associations over a large range of colors. Only knowing the top, or even the top few strongest associated colors with each concept may provide insufficient data for optimal assignment.

One way to quantify human color-concept associations is with human judgments, but collecting such data requires time and effort. A more efficient alternative is to automatically estimate color-concept associations using image or language databases. Previous studies laid important groundwork for how to do so as part of end-to-end methods for palette design \cite{lindner2012a, lindner2012b, setlur2016, havasi2010, jahanian2017}. However, without directly comparing estimated color-concept associations to human judgments, it is unclear how well they match. Further, questions remain about how best to extract information from these databases to match human judgments.

The goal of our study was to understand how to effectively and efficiently estimate  color-concept associations that match human judgments. These estimates can serve as input for palette design, both for creating visualizations and for creating stimuli to use for visual reasoning studies on how people interpret visualizations. 

\parsec{Contributions.} Our main contribution is a new method for automatically estimating color-concept associations in a way that strongly correlates with human ratings. Our method operates directly on Google Image search results, without the need for humans in the loop. Creating an accurate model requires fine-tuning the way in which color information is extracted from images. We found that color extraction was most effective when it used features aligned with perceptual dimensions in color space and cognitive representations of color categories.

To test the different extraction methods, we used a systematic approach starting with simple geometry in color space and building toward methods more grounded in color perception and cognition. We used cross-validation with a set of human ratings to train and validate the model. Once generated, the model can be used to estimate associations for concepts and colors not seen in the training process without requiring additional human data. We demonstrated the effectiveness of this process by training the model using human association ratings between 12 fruit concepts and 58 colors, and testing it on a dataset of 6 recycling-themed concepts and 37 different colors.

\section{Related work}
Several factors are relevant when designing color palettes for visualizing categorical information. First and foremost, colors that represent different categories must appear different; \textit{perceptually discriminable} \cite{healey1996, stone2014, szafir2018, brewer1994}. Other considerations include selecting colors that have distinct names \cite{heer2012}, are aesthetically preferable \cite{gramazio2017}, or evoke particular emotions \cite{bartram2017}. Most relevant to the present work, it is desirable to select \textit{semantically interpretable} color palettes to help people interpret the meanings of colors in visualizations \cite{schloss2018, lin2013, setlur2016}. This can be achieved by selecting ``semantically resonant'' colors, which are colors that evoke particular concepts \cite{lin2013}. It can also be achieved when only a subset of the colors are semantically resonant if conditions support people's ability to infer the other assignments \cite{schloss2018}, see Section~\ref{sec:interpreting}.

\subsection{Creating semantically interpretable color palettes}
Many approaches exist for creating semantically interpretable color palettes  \cite{schloss2018, lin2013, setlur2016}, but they generally involve the same two stages:

\vspace{-3mm}
\begin{enumerate}
\itemsep=-1pt
  \item Quantifying color-concept associations.
  \item Assigning colors to concepts in visualizations, using the color-concept associations from stage 1.
\end{enumerate}
\vspace{-3mm}
Assigning colors to concepts in visualizations (stage~2) relies on input from color-concept associations (stage~1), which suggests assignments are only as good as the association data used to generate them. Color-concept association data are good when they match human judgments. 

\subsubsection{Quantifying color-concept associations}\label{sec:est_color_concept_assoc}
A direct way to quantify human color-concept associations is with humans judgments. Methods include having participants rate association strengths between colors and concepts \cite{ osgood1957, schloss2018}, select colors that best fit concepts \cite{ou2004, dandrade1974, wright1962}, or name concepts associated with colors \cite{palmer2010, munroe2010}. However, collecting these data is time- and resource-intensive.

An alternative approach is to estimate human color-concept associations from large databases, such as tagged images \cite{lin2013,lindner2012a,lindner2012b,setlur2016,bartram2017}, color naming data sets \cite{havasi2010, setlur2016, lindner2012a}, semantic networks \cite{havasi2010}, and natural language corpora \cite{setlur2016}. Each type of database enables linking colors with concepts but has strengths and weaknesses, so automated methods often combine information from multiple databases \cite{lindner2012a, lindner2012b, setlur2016, havasi2010}. 

Extracting colors from tagged images (e.g., Flickr, Google Images) provides detailed color information for a given concept because of the large range of colors within images\cite{lindner2012a, lindner2012b, lin2013, setlur2016, bartram2017}. Methods must specify how to represent colors from images and what parts of the image to include. Linder et~al. \cite{lindner2012a, lindner2012b} obtained a color histogram with bin sizes of $15 \times 15 \times 15$ units in CIELAB space from all pixels in an image. Similarly, Lin et~al. \cite{lin2013} calculated color histograms using smaller $5 \times 5 \times 5$ bins, together with heuristics to smooth the histogram and remove black or white backgrounds. In a different approach, Setlur and Stone \cite{setlur2016} and Bartram et~al. \cite{bartram2017} used clustering algorithms to determine dominant colors in images. Clustering can be effective for finding the top colors in an image, but does not provide information on the full color range. 

Image extraction methods must also specify how to query image databases to obtain images for each concept. Lin et~al. \cite{lin2013} used queries of each concept word alone (e.g., ``apple'')  and queries with ``clipart'' appended to the concept word (e.g.,``apple clipart''). They reasoned that for some concepts, humanmade illustrations would better capture people's associations (e.g., associations between ``money'' and green might be missed from photographs of US dollars, which are more grayish than green). Setlur and Stone \cite{setlur2016} also used ``clipart'', and filtered by color using Google's dominant color filter. 

Language-based databases provide a different approach, linking colors and concepts through naming databases (XKCD color survey \cite{munroe2010} used in \cite{havasi2010, lindner2012b, setlur2016}), concept networks (ConceptNet \cite{liu2004, speer2017} used in \cite{havasi2010}), or linguistic corpora (Google Ngram viewer used in \cite{setlur2016}). Naming databases provide information about color-concept pairs that were spontaneously named by participants. These data can be sparse if they they lack information about concepts that are moderately associated with a color but not strongly associated enough to elicit a color name. Concept networks and linguistic corpora link concepts to color words (not coordinates in a color space), so these methods tend to be used in conjunction with naming \cite{havasi2010} and image \cite{setlur2016} databases to link color words to color coordinates for use in visualizations.

\begin{figure}[tb]
	\centering 
	\includegraphics[width=0.95\columnwidth]{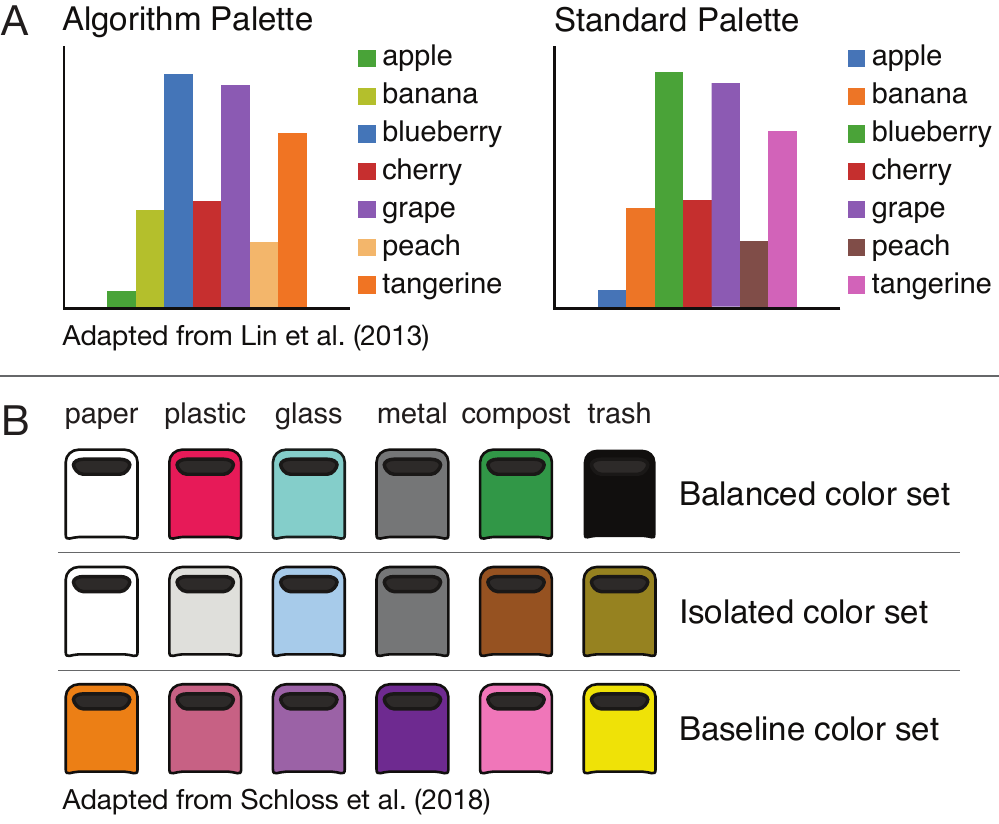}
	\figskip\caption{Examples of designs that use color-concept associations to automatically assign colors to concepts. In (A), data visualizations showed fictitious fruit sales and participants interpreted the colors using legends \cite{lin2013}. In (B), visualizations showed recycling bins and participants interpreted colors without legends or labels \cite{schloss2018}.}
	\figskip\label{fig:CategoricalPalettes}
\end{figure}

\begin{figure*}[ht!]
	\centering 
	\includegraphics[width=0.95\textwidth]{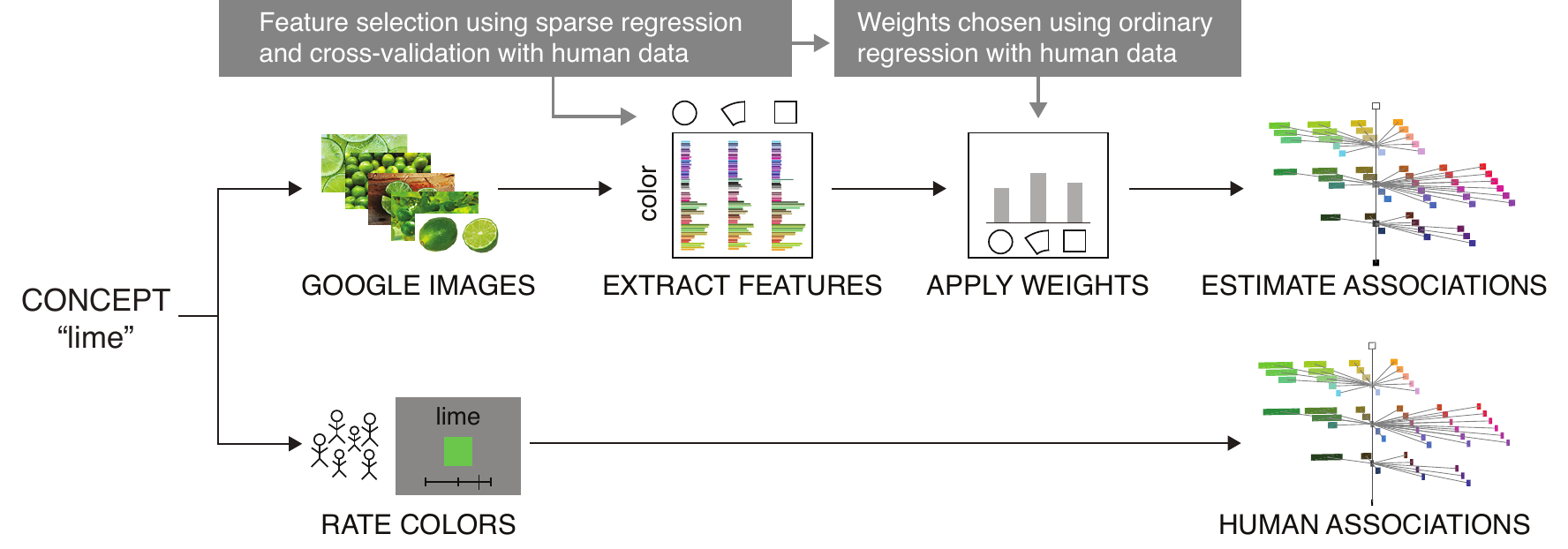}
	\figskip\caption{Illustration of our pipeline for automatically extracting color distributions from images. The bottom flow (concepts to color ratings to human associations) describes the slow yet reliable direct approach using human experiments to determine ground-truth associations. The top flow involves querying Google Images, extracting colors using a variety of different methods (\textit{features}), then weighting those features appropriately to obtain estimated associations. Deciding which features to use and how to weight them is learned from human association data using sparse regression and cross-validation. Once the model is trained, color-concept associations can be quickly estimated for new concepts without additional human data.}
	\figskip\label{fig:Pipeline}
\end{figure*}

\subsubsection{Assigning colors to concepts in visualizations} \label{sec:assigning_colors}
Once color-concept associations are quantified, they can be used to compute color-concept assignments for visualizations. Various methods have been explored for computing assignments. Lin et al. \cite{lin2013} computed \textit{affinity scores} for each color-concept pair using an entropy-based metric, and then solved the assignment problem (a linear program) to find the color-concept assignment that maximized the sum of affinity scores. They found that participants were better at interpreting charts of fictitious fruit sales with color palettes generated from their algorithm, compared with the Tableau 10 standard order (Fig.~\ref{fig:CategoricalPalettes}A). In another approach, Setlur and Stone \cite{setlur2016} applied \textit{k}-means clustering to quantize input colors into visually discriminable clusters using CIELAB Euclidean distance, and iteratively reassigned colors until all color-concept conflicts were resolved. 

In a third approach, Schloss et~al. \cite{schloss2018} solved an assignment problem similar to~\cite{lin2013}, except they computed merit functions (affinity scores) differently. They compared three merit functions: (1) \textit{isolated}, assigning each concept to its most associated color while avoiding conflicts, (2) \textit{balanced}, maximizing association strength while minimizing confusability, and (3) \textit{baseline}, maximizing confusability (Fig.~\ref{fig:CategoricalPalettes}B). They found that participants were able to accurately interpret color meanings for unlabeled recycling bins using the balanced color set, were less accurate for bins using the isolated color set, and were at chance for bins using the baseline color set.

\subsection{Interpreting visualizations of categorical information, and implications for palette design} \label{sec:interpreting}
When people interpret the meanings of colors in visualizations, they use a process called \textit{assignment inference} \cite{schloss2018}. In assignment inference, people infer assignments between colors and concepts that would optimize the association strengths over all paired colors and concepts. Sometimes, that means inferring that concepts are assigned to their strongest associated color (e.g., paper to white in Fig.~\ref{fig:CategoricalPalettes}B). But, other times it means inferring that concepts are assigned to weaker associates, even when there are stronger associates in the visualization (e.g., plastic to red and glass to blue-green, even though both concepts are more strongly associated with white) \cite{schloss2018}. This implies that people can interpret meanings of colors that are not semantically resonant (i.e., strongly associated with the concepts they represent) if there is sufficient context to solve the assignment problem (though what constitutes ``sufficient context'' is the subject of ongoing research). 

People's capacity for assignment inference suggests that for any set of concepts, it is possible to construct many semantically interpretable color palettes. This flexibility will enable designers to navigate trade-offs between semantics and other design objectives---discriminabilty, nameablity, aesthetics, and emotional connotation, which are sometimes conflicting \cite{gramazio2017, bartram2017}. To create palette designs that account for these objectives, it is necessary to have good estimates of human color-concept associations over a broad range of colors.

\section{General method}

In this section, we describe our approach to training and testing our algorithm for automatically estimating human color-concept associations (illustrated in Fig.~\ref{fig:Pipeline}). We began by collecting color-concept association data from human participants to use as ground truth. Then, we used Google Images to query each of the concepts and retrieve relevant images. We tested over 180 different methods (\textit{features}) across Experiments 1A-C for extracting color distributions from images.  We selected features (how many and which ones to use) by applying sparse regression with cross-validation to avoid over-fitting and produce estimates that generalized well. Model weights for each feature were chosen by ordinary linear regression. 

For training and testing in Experiments 1 and 2, we used a set of 58 colors uniformly sampled in CIELAB color space ($\Delta E = 25$), which we call the UW-58 colors (see Supplementary Material Section \ref{sec:defining_colors} and Supplementary Table~\ref{table:UW_58_colors}). The concepts were 12 fruits: \textit{avocado}, \textit{blueberry}, \textit{cantaloupe}, \textit{grapefruit}, \textit{honeydew}, \textit{lemon}, \textit{lime}, \textit{mango}, \textit{orange}, \textit{raspberry}, \textit{strawberry}, and \textit{watermelon}. We chose fruits, as in prior work \cite{lin2013, setlur2016}, because fruits have characteristic, directly observable colors (high color diagnosticity \cite{tanaka1999}). We sought to establish our method for simple cases like fruit where we believed there was sufficient color information within images to estimate human color-concept associations. In future work it will be possible to identify edge cases where the method is less effective and address those limitations. In Experiment~3, we tested how our trained algorithm generalized to a different set of colors and concepts using color-concept association ratings for recycling concepts from \cite{schloss2018}.
In the remainder of this section, we describe our methods in detail.

\subsection{Human ratings of color-concept associations} \label{sec:human_ratings}
To obtain ground truth for training and testing our models, we had participants rate association strengths between each of the UW-58 colors and 12 fruits.

\parsec{Participants.} We tested 55 undergraduates (mean age $= 18.52$, 31~females, 24~males) at UW--Madison. Data was missing from one participant (technical error). All had normal color vision (screened with HRR Pseudoisochromatic Plates \cite{hardy2002}), gave informed consent, and received partial course credit. The UW--Madison Internal Review Board approved the protocol.

\begin{figure}[ht!]
	\centering 
	\includegraphics[width=0.75\columnwidth]{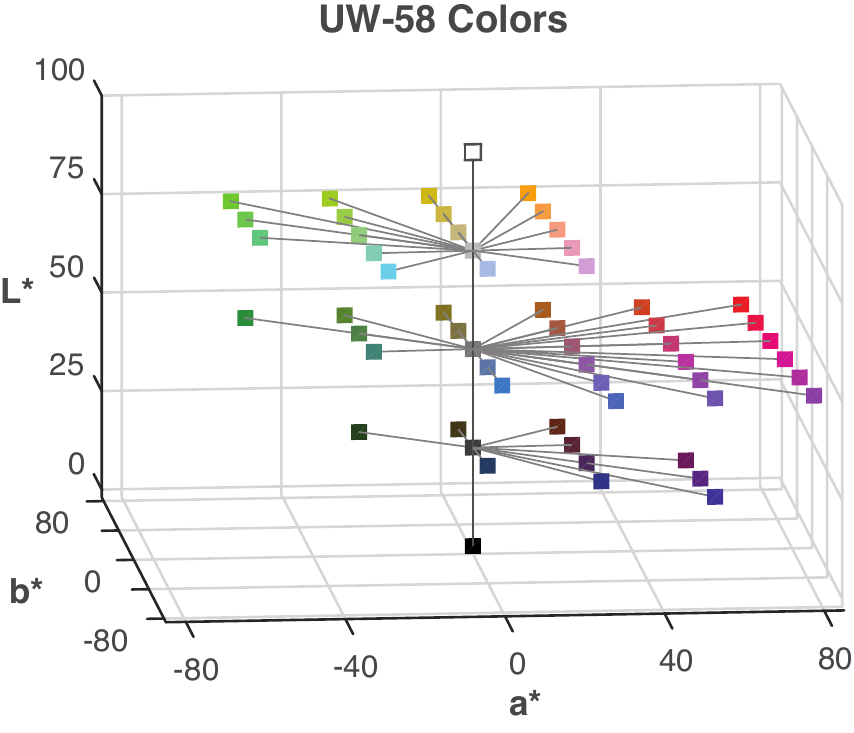}
	\figskip\caption{UW-58 colors used in Experiment 1 and 2, plotted in CIELAB color space. Exact coordinates are given in Supplementary Table~\ref{table:UW_58_colors}.}
	\figskip\label{fig:UW58}
\end{figure}

\parsec{Design, displays, and procedure.}
Participants rated how much they associated each of the UW-58 colors with each of 12 fruit concepts. Displays on each trial contained a concept name at the top of the screen in black text, a colored square centered below the name ($100 \times 100$ pixels), and a line-mark slider scale centered below the colored square. The left end point of the scale was labeled ``not at all'', the right end point labeled ``very much'', and the center point was marked with a vertical divider. Participants made their ratings by sliding the cursor along the response scale and clicking to record their response. Displays remained on the screen until response. Trials were separated by a 500~ms inter-trial interval. All 58 colors were rated for a given concept before going onto the next concept. The order of the concepts and the order of colors within each concept were randomly generated for each participant. 

Before beginning, participants completed an anchoring procedure so they knew what it meant to associate ``not at all'' and ``very much'' in the context of these concepts and colors \cite{palmer2013}. While viewing a display showing all colors and a list of all concepts, they pointed to the color they associated most/least with each concept. They were told to rate those colors near the endpoints of the scale during the task. 

Displays were presented on a 24.1~in ASUS~ProArt~PA249Q monitor ($1920 \times 1200$ resolution), viewed from a distance of about 60~cm. The background was gray (CIE Illuminant D65, x = .3127, y = .3290, Y = $10$~cd/$\text{m}^2$). The task was run using Presentation (\url{www.neurobs.com}). We used a Photo Research PR-655 SpectraScan spectroradiometer to calibrate the monitor and verify accurate presentation of the colors. The deviance between the measured colors and target colors in CIE~1931 xyY coordinates was $< .01$ for x and y, and $< 1$~cd/$\text{m}^2$ for Y.

The mean ratings for all fruit-color pairs averaged over all participants are shown in Supplementary Figs.~\ref{fig:FruitTree1} and~\ref{fig:FruitTree2} and Supplementary Table~\ref{table:human_ratings}.

\subsection{Training color extraction and testing performance}\label{sec:training_extraction}

 Our goal was to learn a method for extracting color distributions from images such that the extracted profiles were reliable estimates of human color-concept association ratings from Section~\ref{sec:human_ratings}. 
 We will describe our approach here using generic parameter names. The parameter values we actually used in the experiments are specified in the relevant experiment descriptions (Section~\ref{sec:exp1}).

For each of the $n_\text{con}$ concepts, we queried Google Images using the name of the concept and downloaded the top $n_\text{img}$ results. We then compiled a list of $n_\text{feat}$ \textit{features}. A feature is a function $f: (\text{image},\text{color}) \to \R$ that quantifies the presence of a given target color in a given image. For example, a feature could be ``the proportion of pixels in the middle 20\% of the image that are within $\Delta r=40$ of the target color''. For each of the $n_\text{img}$ images and each of the $n_\text{col}$ colors, we evaluated each of the $n_\text{feat}$ features. This resulted in a matrix $X \in \R^{n_\text{con}n_\text{img}n_\text{col} \times n_\text{feat}}$, where each row was a $(\text{concept},\text{image},\text{color})$ triplet and each column was a different feature.
We then constructed a vector $y \in \R^{n_\text{con}n_\text{img}n_\text{col} \times 1}$ such that the $i^\text{th}$ element of $y$ contains the average color-concept rating from the human experiments, for the concept and color used in the $i^\text{th}$ row of $X$. Note that each rating in $y$ was repeated $n_\text{img}$ times because for each $(\text{concept},\text{color})$ pair, there are $n_\text{img}$ images.
We used the data $(X,y)$ in two ways: (1) to select how many features to use and (2) to choose feature weights.

\parsec{Feature selection.} We used sparse regression (lasso) with leave-one-out cross-validation to select features. Specifically, we selected a concept, partitioned $(X,y)$ by rows into test data $(X_\text{test},y_\text{test})$, containing the rows pertaining to the selected concept, and training data $(X_\text{train},y_\text{train})$, containing the remaining concepts. We then used sparse regression on the training data with a sweep of the regularization parameter. This was repeated with every concept and we plotted the average test error versus the number of nonzero weights (Fig.~\ref{fig:MSE_select}). We examined the plot and found that $k=4$ features provided a good trade-off between error and model complexity, so we used $k=4$ features for all subsequent experiments. To select which features to use, we ran one final sparse regression using the full data $(X,y)$ and chose the regularization parameter such that $k=4$ features emerged.

Using cross-validation for model selection is standard practice in modern data science workflows. By validating the model against data that was unseen during the training phase, we are protected against overfitting and we help ensure that our model will generalize to unseen concepts and colors or different training images.

\parsec{Choosing feature weights.} Once the best $k$ features were identified, we selected the weights by performing an ordinary linear regression with these $k$ features. We ensured that human data used to compute the model weights were always different from human data used test model performance. In Experiment 1 and 2, when we trained and tested on fruit concepts, we chose feature weights using 11 fruits and tested on the 12th fruit (repeated for each fruit). In Experiment 3, we used a model trained on all 12 fruit concepts to test how well it predicted data for recycling concepts (Section~\ref{sec:exp3}). 

\parsec{Testing the model on new data.} Once the model has been trained, it can be used to estimate color-concept associations for new concepts \textit{and} new colors without the need to gather any new human ratings. The concept is queried in Google Images, the $k$ chosen features are extracted from the images for the desired colors, and the trained model weights are applied to the features to obtain the estimates.

\parsec{Reproducible experiments.} We used cross-validation in order to ensure our results hold more broadly beyond our chosen concepts, colors, and images.
We developed code in Python for all the experiments and made use of the \texttt{scikit-image}~\cite{scikit-image} and \texttt{scikit-learn}~\cite{scikit-learn} libraries to perform all the image processing and regression tasks.
Our code is available at \url{https://github.com/Raginii/Color-Concept-Associations-using-Google-Images}. This repository can be downloaded to a local machine to replicate the entire study or adapt it to test new concepts and colors. The repository also contains a write-up with detailed instructions.

%%%%%%% EXPERIMENT 1 %%%%%%%
\section{Experiments}\label{sec:exp1}
In Experiments 1A--1C, we systematically tested methods for extracting colors from images and assessed model fits with human color-concept association ratings. In Experiment 2, we examined model performance using different image types (top 50 Google Image downloads, cartoons, and photographs). In Experiment 3, we tested how well our best model from Experiment 1 generalized to a different set of concepts and colors. 

\subsection{Experiment 1A: Balls in Cartesian coordinates}
Perhaps the simplest approach for extracting colors from images would be to (1) define a set of target colors of interest in a color space (e.g., CIELAB), (2) query a concept in Google Images (e.g., ``blueberry''), (3) download images returned for that concept, and (4) count the number of pixels in each image that has each target color within the set. However, that level of precision would exclude many colors in images, including some that are perceptually indistinguishable from the target colors. Moreover, not all pixels in the image may be relevant. For example, when people take pictures of objects, they tend to put the object near the center of the frame \cite{palmer2008}, so it is sensible that images ranked highly in Google Images for particular query terms will have the most relevant content near the center of the frame.

In this experiment, we varied \textit{color tolerance}: allowing colors that are not perfect matches with targets to still be counted, and \textit{spatial windows}: including subsets of the pixels that may be more likely to contain relevant colors for the concept. Our goal was to determine which combination of color tolerance and spatial window best captured human color-concept associations.

\subsubsection{Methods} \label{FeatureSelection}
In this section, we explain how we downloaded and processed the images, constructed features, and selected features for our model.

\parsec{Images.}\label{section:Images} We used the same set of testing and training images throughout Experiment 1, downloaded for each fruit using Vasa's \cite{vasa2018} Google Images Download Python script available on GitHub. We used the top $n_\text{img}=50$ images returned for each fruit that were in \texttt{.jpg} format, but our results are robust to using a different number of images (see Supplementary Section~\ref{sec:varying_number_of_images}). We did not use the Google Image Search API used in prior work~\cite{lin2013, setlur2016} because it has since been deprecated. We made the images uniform size by re-scaling them to $100 \times 100$ pixels, which often changed the aspect ratio.
 
We converted RGB values in the images to CIELAB coordinates using the \texttt{rgb2lab} function in \texttt{scikit-image}. This conversion makes assumptions about the monitor white point and only approximates true CIELAB coordinates. It is standard to use this approximation in visualization research given that in practice, people view visualizations on unstandardized monitors under uncontrolled conditions \cite{heer2012, szafir2018, stone2014, gramazio2017}. Although our model estimations are constructed based on approximations of CIELAB coordinates, our human data were collected on a calibrated monitor with true CIELAB coordinates (see Section~\ref{sec:human_ratings}). Thus, the fit between human ratings and model approximations speaks to the robustness of our approach.

\parsec{Features.} We constructed 30 features from all possible combinations of 5 color tolerances and 6 spatial windows as described below.

\textit{Color tolerances.}  When looking for a target color within a set of pixels in an image (set defined by the spatial window), we counted the fraction of all pixels under consideration whose color fell within a ball of radius $\Delta r$ in CIELAB space of the target color. We tested balls with five possible values of $\Delta r$: 1, 10, 20, 30, and 40. 

\textit{Spatial windows.} We varied spatial windows in six ways. The first five extracted pixels from the center 20\%, 40\%, 60\%, 80\%, and 100\% of the image, measured as a proportion of the total area.   The $6^\text{th}$ way used a figure-ground segmentation algorithm to select figural, ``object-like'' regions from the image, and exclude the background. Here ``window'' is not rectangular, but rather the shape of the figural region(s) as estimated by the active contour algorithm \cite{michael1998snakes,whitaker1998level}, \textit{Snakes}. This uses an initial contour binary mask and iteratively moves to find the object boundaries. We used the MATLAB implementation \texttt{activecontour} from the Image Processing Toolbox for 500 iterations, setting the initial contour as the image boundary. Although prior work suggested that figure-ground segmentation might not provide a benefit beyond eliminating borders \cite{lin2013}, we aimed to test its effects in the context of our other color tolerance and spatial window parameters.

\begin{figure}[tb]
	\centering
	\includegraphics[width=0.85\columnwidth]{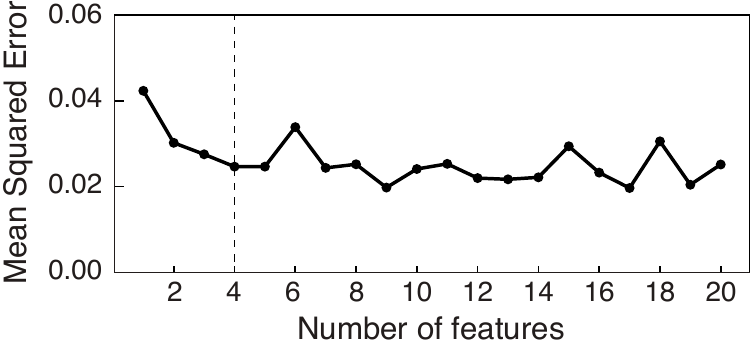}
	\figskip\caption{Mean squared test error (MSE) as a function of the number of features selected using sparse regression (lasso). MSE is averaged over 12 models obtained using leave-one-out cross-validation, using each fruit category as different test set. We selected models with four features.}
	\figskip\label{fig:MSE_select}
\end{figure}

\parsec{Feature selection.} We applied the feature selection method described in Section~\ref{sec:training_extraction} using the 30 features described above, as well as a constant offset term, which is standard when using regression. Using an offset is equivalent to adding one more feature equal to the constant function 1. Fig.~\ref{fig:MSE_select} shows the mean squared error (MSE) for each number of features,  averaged across all $n_\text{con}=12$ fruits, using the $n_\text{col}=58$ UW-58 colors, and using $n_\text{img}=50$ Google Image query results for each fruit.

Based on this plot, we decided to use 4 features, which yields a good trade-off between model complexity and error reduction. We then used the full dataset to select the best 4 features, as described in Section~\ref{sec:training_extraction}. 

The best 4 features were (1) constant offset, (2) 20\% window with $\Delta r=40$ (positive weight), (3) 100\% window with $\Delta r=40$ (negative weight), and (4) segmented figure with $\Delta r=40$ (positive weight), see Table~\ref{table:WeightsForAllModels}. The positive weight on the center of the image and negative weight on 100\% of the image can be interpreted as a crude form of background suppression. 

\begin{table}[ht]
\caption{Top 4 features selected using sparse regression as more features were made available in Experiments 1A to 1C. Ball features were not selected when sector or category features became available.}\label{table:WeightsForAllModels} 
\centering
\begin{tabular}{cl} 
\toprule
\textbf{Model description}   & \textbf{Features selected} \\ \midrule
\textbf{Ball model}& constant offset \\
(Experiment 1A)                & Ball: $\Delta r=40$; $20\%$ window\\    
Features available:     & Ball: $\Delta r=40$; $100\%$ window\\           
Ball only & Ball: $\Delta r=40$; segmented\\
\midrule
\textbf{Sector model}& constant offset \\
(Experiment 1B)                    & Sector: $\Delta r=40$, $\Delta h=40\degree$; $20\%$ window\\
Features available:         & Sector: $\Delta r=40$, $\Delta h=30\degree$; $40\%$ window\\
Ball, Sector     & Sector: $\Delta r=40$, $\Delta h=40\degree$; segmented\\
\midrule
\textbf{Sector+Cat model} & constant offset \\
(Experiment 1C)                    & Sector: $\Delta r=40$, $\Delta h=40\degree$; $20\%$ window\\ 
Features available: & Sector: $\Delta r=40$, $\Delta h=40\degree$; segmented\\
Ball, Sector, Category     & Category; $20\%$ window\\
\bottomrule 
\end{tabular}
\figskip
\end{table}

\subsubsection{Results and discussion}
\begin{figure*}[tb]
	\centering
	\includegraphics[width=\textwidth]{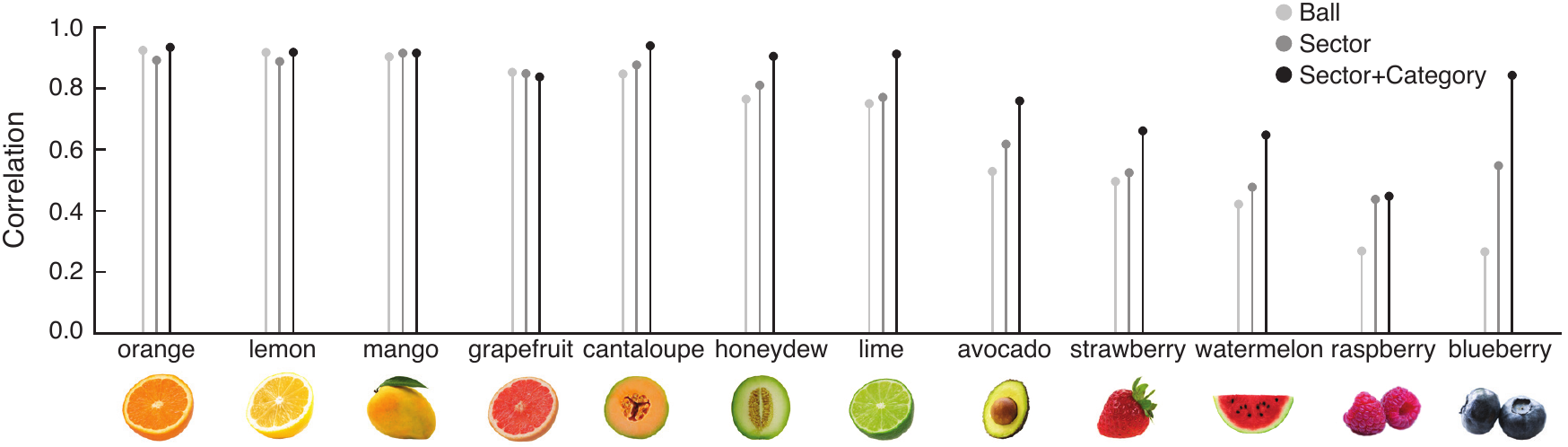}
	\figskip\figskip\caption{Correlations between model estimations and human ratings across each of the UW-58 colors for each fruit from the best 4-feature model in Experiment~1A (Ball model),  Experiment~1B (Sector model), and Experiment~1C (Sector+Category model). The Sector+Category model performed best, followed by the Sector model, then the Ball model, see Table \ref{table:Exp1Corrs} and text for statistics.}
	\figskip\label{fig:E1Corr}
\end{figure*}

We tested the model on each of the fruits using the leave-one-out cross-validation procedure described in Section~\ref{sec:training_extraction}. We trained model weights using the $11\times 50 = 550$ training images and averaged the model estimates across the $50$ test images. We tested the effectiveness of the Ball model by correlating its estimates with mean human ratings over all 12 fruits $\times$ 58 colors and found a moderate correlation of .65 (Table \ref{table:Exp1Corrs}). Fig.~\ref{fig:E1Corr} shows the correlations separately for each fruit (light gray points), with fruits sorted along the $x$-axis from highest to lowest $r$ value for the Ball model. There is wide variability in the model fits, ranging from $r=.93$ for \textit{orange} to $r=.27$ for \textit{blueberry}. 

\begin{figure}[tb]
	\centering
	\includegraphics[width=0.9\columnwidth]{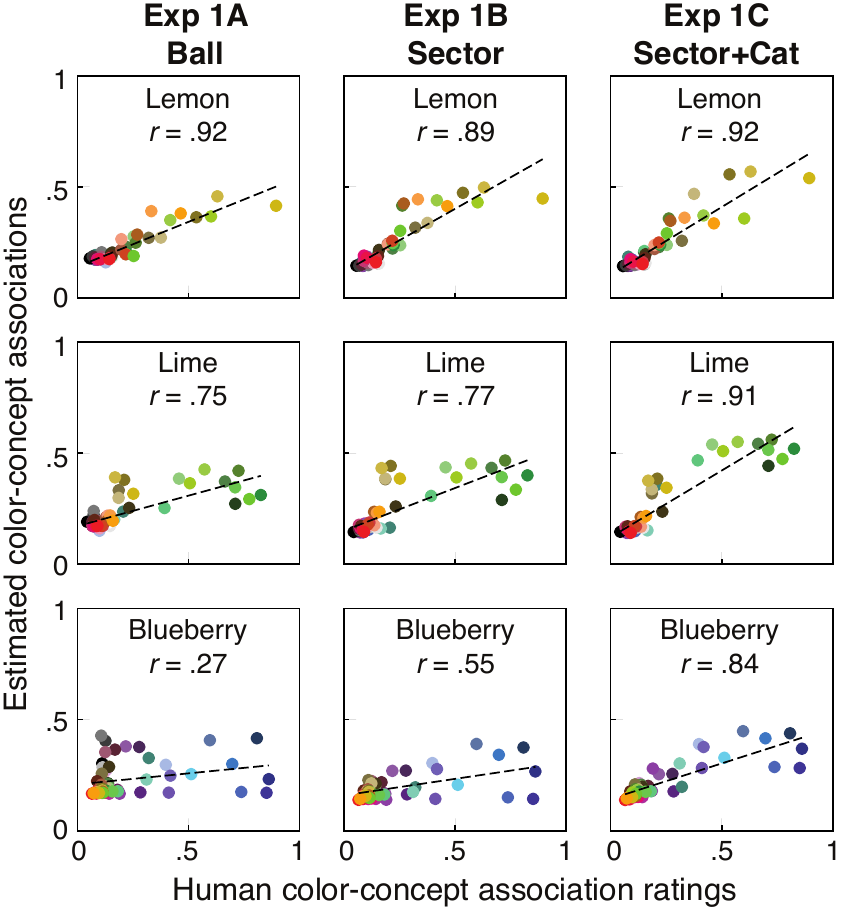}
	\figskip\caption{Scatter plots showing relationships between model estimates and human ratings for \textit{lemon}, \textit{lime}, and \textit{blueberry} using models from Experiments 1A--1C. Marks represent each of the UW-58 colors, dashed line represent best-fit regression lines, and \textit{r} values indicate correlations within each plot. Adding more perceptually relevant (sector) and cognitively relevant (category) features improved fit for fruits where ball features performed poorly.}
	\figskip\label{fig:E1Scatters}
\end{figure}

To understand why the model performed poorly for some fruits, we plotted estimated ratings for each color as a function of human ratings. Fig.~\ref{fig:E1Scatters} highlights a subset of three fruits with high, medium, and low correlations. The full set of plots are shown in Supplementary Fig.~\ref{fig:AllScatter1}. Error in performance seemed to arise from underestimating the association strength for colors that did not appear in the images but were associated with the concepts. This was particularly apparent for \textit{blueberry}, where participants strongly associated a variety of blues that were more saturated and purplish than the blues that appeared in images of blueberries. Model estimates for those blues were as low as model estimates for oranges and greens, which were clearly \textit{not} associated with \textit{blueberry}. The model also overestimated values for grays and purples that were not associated with \textit{blueberry}. These results suggested that different kinds of features would be necessary for capturing human color-concept associations.

\begin{table}[ht]
\caption{Correlations ($r$) between mean human color-concept association ratings and estimated associations (12 fruits $\times$ 58 colors $=$ 696 items) for each model in Experiments 1 and 2 (also shown Fig.\ref{fig:E2CorrErr}). All $p$ $<.001$.}\label{table:Exp1Corrs} 
\centering
\begin{tabular}{lccc} 
\toprule
\textbf{Model} & \textbf{Top 50} & \textbf{Photo}  & \textbf{Cartoon}\\ \midrule
Ball                & .65 &  .62  & .72\\
Sector              & .72 &  .69  & .72 \\
 Sector+Category    & .81 &  .80  & .80\\
\bottomrule 
\end{tabular}
\figskip
\end{table}

\subsection{Experiment 1B: Sectors in cylindrical coordinates}
A potential limitation of the ball features in Experiment~1A is that varying the size of the ball has different perceptual consequences depending on the location in color space. This is because perceptual dimensions of color are cylindrical (angle: hue, radius: chroma, height: lightness) rather than Cartesian. Balls of a fixed size that are closer to the central L* axis will span a greater range of hue angles than balls that are farther away (i.e., higher chroma). In the extreme, a ball centered on a*~$=0$ and b*~$=0$ (e.g., a shade of gray) will include colors of all hues. Thus, if we wanted a ball that was large enough to subsume all the high chroma blues (i.e., colors strongly associated with blueberries), that same large ball placed near the achromatic axis would subsume all hues of sizable chroma (see Figure \ref{fig:CirclesSectors}). 

To have independent control over hue and chroma variability, we defined new features with tolerance regions as cylindrical sectors around the target colors according to hue angle and chroma (Fig.~\ref{fig:CirclesSectors}).  We tested whether color extraction using sector features, more aligned with perceptual dimensions of color space, would better estimate human color-concept associations. We used the same images as in Experiment~1A, and confined the number of features to four, using the same sparse regression approach as in Experiment~1A for feature selection. We assessed whether the best model included any of these new features, and if so, if that model's estimates fit human ratings significantly better than the Ball model did in Experiment~1A. 

\subsubsection{Methods}
We included the same 30 features from Experiment 1A, plus 150 new features: 25 new color tolerance regions $\times$ 6 spatial windows (same spatial windows as Experiment 1A) for a total of 180 features. The 25 new color tolerance regions were defined in cylindrical coordinates in CIELch space using all combinations of 5 hue angle tolerances ($\Delta h$: 5\degree, 10\degree, 20\degree, 30\degree, 40\degree) and 5 chroma/lightness tolerances ($\Delta r$: 1, 10, 20, 30, 40) around each target color, see Fig.~\ref{fig:CirclesSectors}. The tolerances for chroma and lightness co-varied, so $\Delta r = 10$ means that both chroma and lightness had a tolerance of 10. Note that CIELch space is the same as CIELAB space except it uses cylindrical rather than Cartesian coordinates.

\subsubsection{Results and discussion}
As in Section \ref{FeatureSelection}, we first extracted the best 4 features from the pool of 180 features using sparse regression. The 4 top features only included sector features and no ball features (Table~\ref{table:WeightsForAllModels}), so we refer to the model from Experiment 1B as the ``Sector'' model.

We tested the effectiveness of the Sector model by correlating its estimates with mean human ratings over all 12 fruits $\times$ 58 colors. This correlation (\textit{r} = .72) was stronger for the Sector model than for the Ball model (Table \ref{table:Exp1Corrs}), and that difference was significant ($z = 2.46$, $p = .014$). Fig.~\ref{fig:E1Corr} (medium gray points) shows that the Sector model improved performance for fruits that had the weakest correlations in Experiment~1A, but there is still room for improvement. The scatter plots in Fig.~\ref{fig:E1Scatters} show that the model still under-predicts ratings for blues that are strongly associated with blueberries but are not found in the images. A similar problem can be observed for limes, where several greens are associated with limes, but are not extracted from the images. The full set of scatter plots is in Supplementary Fig.~\ref{fig:AllScatter2}. 

These results suggest that when people form color-concept associations, they might extrapolate to colors that are not directly observed from visual input. We address this possibility in Experiment~1C. 

\subsection{Experiment 1C: Color categories} \label{sec:exp_color_categories}
 In Experiment 1C, we examined the possibility that human color-concept associations extrapolate to colors that are not directly observed from visual input. One way that extrapolation might occur is through color categorization. Although colors exist in a continuous space, humans partition this space into discrete categories. English speakers use 11 color categories with the basic color terms red, green, blue, yellow, black, white, gray, orange, purple, brown, and pink \cite{berlin1969}. The number of basic color terms varies across languages \cite{berlin1969, gibson2017, roberson2005, kay2003, witzel2018}, but there are regularities in the locus of categories in color space \cite{kay2003,abbott2016}. 

We propose that when people form color-concept associations from visual input, they extrapolate to other colors that are not in the visual input but share the same category (\textit{category extrapolation hypothesis)}. To test this hypothesis, it was necessary to first identify the categories of each color within images. We did so using a method provided by Parraga and Akbarinia \cite{parraga2016}, which used psychophysical data to determine category boundaries for each of the 11 color terms. Their algorithm enables efficient lookup and categorization of each pixel within and image.
We then constructed a new type of feature that represents the proportion of pixels in the image that share the color category of each of the UW-58 colors. For example, if .60 of the pixels in the image are categorized as ``blue'' (using \cite{parraga2016}), then all UW-58 colors that are also categorized as ``blue'' will receive a feature value of .60, regardless of how much of those UW-58 colors were in the image. We assessed whether the best model included any category new features, and if so, whether that model's estimates fit human rating significantly better than the Sector model did in Experiment~1B.

\subsubsection{Methods}

We included the 30 ball and 150 sector features in Experiments~1A and~1B, plus 6 new color category features for a total of 186 features. We generated category features using Parraga and Akbarinia's \cite{parraga2016} method to obtain the color categories of our UW-58 colors and the categories of each pixel within our images. We used the functions available through their Github repository \cite{ColourCategorisation2017} to convert RGB color coordinates to color categories. Specifically, we converted the $100\times 100\times 3$ image arrays in RGB to $100 \times 100$ arrays, where each element in the array represents the pixel's color category. For each UW-58 color, we defined the features to be the fraction of pixels in the spatial window that belonged to the same color category as the UW-58 color. We repeated the above procedure with the 6 spatial windows as before.  

\subsubsection{Results and discussion}
Similar to Experiments 1A and 1B, we used sparse regression to extract the best 4 features among 186 total features. The model selected the constant offset and two of the same sector features from Experiment~1B, plus one new category feature (no ball features), see Table~\ref{table:WeightsForAllModels}). Thus, we refer to this new model as the ``Sector+Category'' model. We obtained the model weights via linear regression as detailed in Section~\ref{sec:training_extraction}.

We tested the effectiveness of the Sector+Category model by correlating its estimates with mean human ratings over all 12 fruits $\times$ 58 colors. This correlation was stronger for the Sector+Category model than for the Ball model or Sector model (Table \ref{table:Exp1Corrs}), and those differences were significant ($z=6.55$, $p<.001$; $z=4.08$, $p<.001$, respectively). Fig.~\ref{fig:E1Corr} shows that the Sector+Category model (black points) further improved fits for the fruits that had weaker fits using the other two models. As seen in Fig.~\ref{fig:E1Scatters}, by including category extrapolation, this model increased the estimated values for colors that were strongly associated with limes and blueberries (greens and blues, respectively) that were under-predicted by the previous models because those colors were not in the images. The full set of scatter plots is in Supplementary Fig.~\ref{fig:AllScatter3}.

%%%%%%% EXPERIMENT 2 %%%%%%%
\subsection{Experiment 2: Comparing image types}\label{sec:exp2}
As described in Section~\ref{sec:est_color_concept_assoc}, Lin et~al. \cite{lin2013} queried concept words alone and concept words appended with ``clipart'', reasoning that humanmade illustrations might better capture associations for some types of concepts. We propose that humans produce clipart illustrations based on their color-concept associations, not solely based on real-world color input. If color-concept associations are already incorporated into clipart, that would explain why clipart is useful for estimating color-concept associations, especially when natural images fall short. However, if a model contains features that effectively estimate human color-concept associations, it may have sufficient information to do as well for natural images as it does for humanmade illustrations, such as clipart. To examine this hypothesis, we tested our models from Experiments 1A-C on two new image types: cartoons (humanmade illustrations) and photographs (not illustrations). 

In addition, we used the approach of Lin~et.~al. (mentioned in Section~\ref{sec:assigning_colors}) to compute probabilities, a precursor to their affinity score that most corresponds to color-concept associations. We then compared those probabilities with our human ratings.

\subsubsection{Methods}
 We downloaded two new sets of images, by querying each fruit name appended with ``cartoon'' or ``photo''.  We queried ``cartoon'' rather than ``clipart'' because clipart sometimes contained parts of photographs with the background deleted, and we wanted to constraint this image set to humanmade illustrations. Unlike Experiment~1 where we used the first 50 images returned by Google Images, we manually curated the photo and cartoon image sets to ensure (a) they were photos for the photo set and cartoons in the cartoon set, (b) they included an observable image of the queried fruit somewhere in the image, and (c) they were not images of cartoon characters (e.g., ``Strawberry Shortcake'', a character in an animated children's TV show that first aired in 2003). This resulted in 50 images in each set.

We trained and tested using the same top 4 features from the Ball, Sector, and Sector+Category models in Experiment~1, except we substituted the training images for the manually curated sets of cartoon images or photo images. This yielded three sets of model weights for the different image types. We then compared each model's performance for the three image types.

\subsubsection{Results and discussion}

\begin{figure}[b]
    \centering 
	\includegraphics[width=0.85\columnwidth]{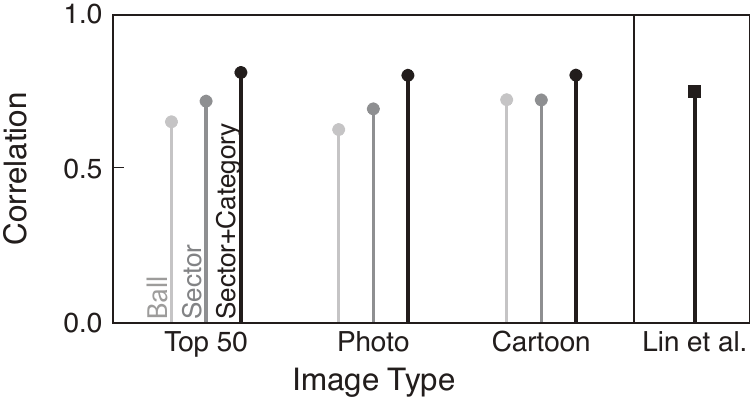}
	\figskip\caption{Correlations for top 50 images, photo images, and cartoon images using the Ball, Sector, and Sector+Category models. The Sector+Category model performed best and was similar across all image types. The Ball model was worst for top 50 and photo images, but less poor for cartoon images.  Estimates based on Lin et al. \cite{lin2013} were strongly correlated with human ratings, but less so than the Sector+Category model (see text for statistics and explanation).}
	\label{fig:E2CorrErr}
\end{figure}

Fig.~\ref{fig:E2CorrErr} and Table \ref{table:Exp1Corrs} show the overall correlations between color-concept associations and model estimates across all 12 fruits $\times$ 58 colors. The correlations for the top 50 images are those previously reported in Experiment~1A--1C, and included here as a baseline. Fig.~\ref{fig:E2CorrErr} also shows overall correlations with the probabilities computed using the method of Lin~et.~al.~\cite{lin2013} as another baseline. 

The results suggest there is a benefit to using human-illustrated cartoons for the Ball model (which does not effectively capture human color-concept associations), but the benefit diminishes for the Sector and Sector+Category models (which better capture human color-concept associations). 
Specifically, the Ball model using cartoons was significantly more correlated with human ratings than the Ball model using top 50 images ($z=2.46$, $p=.014$) with no difference between cartoon and top 50 images for the other two models (Sector: $z=0.0$, $p=1.0$; Sector+Category: $z=.53$, $p=.596$). There were no significant differences in fits using photo vs. top 50 images for any of the three model types (Ball: $z=0.94$, $p=.347$; Sector: $z=1.11$, $p=.267$; Sector+Category: $z=0.53$, $p=.596$). 

To further understand these differences, we tested for interactions between image type and feature type, using linear mixed-effect regression (R version 3.4.1, lme4 1.1-13, see \cite{brauer2017}). The dependent measure was model error for each fruit (sum of the squared errors across colors for each fruit). We included fixed effects for Model, Image, and their interactions, and random slopes and intercepts for fruit type within each Model contrast and Image contrast. We initially tried including random slopes and intercepts for  interactions, but the model became too large and the solver did not converge. We tested two contrasts for the Model factor. The first was Category+Sector vs. average of Ball and Sector, which enabled us to test whether Category+Sector was overall better than the other two models. The second was Sector vs.\ Ball, which enabled us to test whether the Sector model was better than the Ball model. We tested two contrasts for the Image factor. The first was cartoon vs.\ average of top 50 and photo, which enabled us to test whether cartoons were overall better than the other two images types. The second contrast was top 50 vs.\ photo, which enabled us to test whether top 50 images (which included some cartoons) were better than photos. Reported beta and t-values are absolute values.

The results for the Model contrasts showed that Sector+Category model preformed best, and the Sector model performed better than the Ball model. That is, there was less error for  Sector+Category than the combination of Ball and Sector ($\beta= 0.10$, $t(11) = 4.42$, $p=.001$), and less error for Sector than for Ball ($\beta=0.083$, $t(11)=7.90$, $p<.001$). The contrasts for Image were not significant ($ts<1$), indicating no overall benefit for human made cartoons. 

However, the first Image contrast comparing cartoons vs. the average of top 50 and photo interacted with both Model contrasts. Looking at the interaction with the first Model contrast (Sector+Category vs. average of Ball and Sector), the degree to which Sector+Category model outperformed the other models was greater for top 50 and photo images than for cartoons ($\beta=0.01$, $t(44)=4.38$, $p<.001$), see Fig. \ref{fig:E2CorrErr}. Looking at the interaction with the second Model contrast (Sector vs. Ball), the degree to which Sector model outperformed the Ball model was greater for the top 50 and photo images than the cartoons ($\beta=0.02$, $t(44)=6.65$, $p<.001$). No other interactions were significant ($ts< 1$).

In this experiment, we also evaluate how Lin et al.'s estimates of color-concept associations \cite{lin2013} match our human ratings. Their estimates come from a hybrid of color distributions extracted from top image downloads and clipart, so we provide their model with our top 50 images and cartoons as input.  As shown in Fig.~\ref{fig:E2CorrErr}, the correlation for all fruits and colors was $r=.74$, which is similar to our Sector models ($r=.69$ to .72 depending on image type) and less strong than our Sector+Category models ($r=.80$ to .81 depending on image type). The difference in correlation for the Lin et al. model and our Sector+Category model for the top 50 images was significant ($z=3.29$, $p<.001$). We note that these models are not directly comparable because our models used either top 50 images or cartoons, not both at the same time (except if cartoons happened to appear in the top 50 images).

In summary, using cartoon images instead of other image types helped compensate for the poor performance of our Ball model. However, image type made no difference for our more effective Sector and Sector+Category models. We interpret these results as showing that cartoons help the Ball model compensate for poorer performance because humans make cartoons in a way that builds in aspects of human color-concept associations. For example, visual inspection suggests that cartoon blueberries tend to contain saturated blues that are highly associated with blueberries yet not present in photographs of blueberries. However, the benefit of humanmade illustrations is reduced if model features are better able to capture human color-concept associations. This suggests that using our Sector+Category models on the top Google Image downloads is sufficient for estimating human color-concept associations without further need to curate the image set, at least for the concepts tested here.

%%%%%%% EXPERIMENT 3 %%%%%%%
\subsection{Experiment 3: Generalizing beyond fruit}\label{sec:exp3}
Experiment 3 tested how well our model trained on fruit generalized to a new set of concepts and colors using the recycling color-concept association dataset from \cite{schloss2018}. The concepts were: \textit{paper}, \textit{plastic}, \textit{glass}, \textit{metal}, \textit{compost}, and \textit{trash}, and colors were the BCP-37 (see Supplementary Table~\ref{table:BCP_37_colors}). Unlike fruits, which have characteristic colors, recyclables and trash can come in any color. Still, human ratings show systematic color-concept associations for these concepts, and we aimed to see how well our model could estimate those ratings. 
 
\subsubsection{Methods} 
As in Experiment 1, we downloaded the top 50 images from Google Images for each recycling-related concept. To estimate color-concept associations, we used our Sector+Category model from Experiment 1C with feature weights determined from a single linear regression using all 12 fruits, as described in Section~\ref{sec:exp_color_categories}.

\subsubsection{Results and discussion}
\begin{figure}[tb]
	\centering 
	\includegraphics[width=0.85\columnwidth]{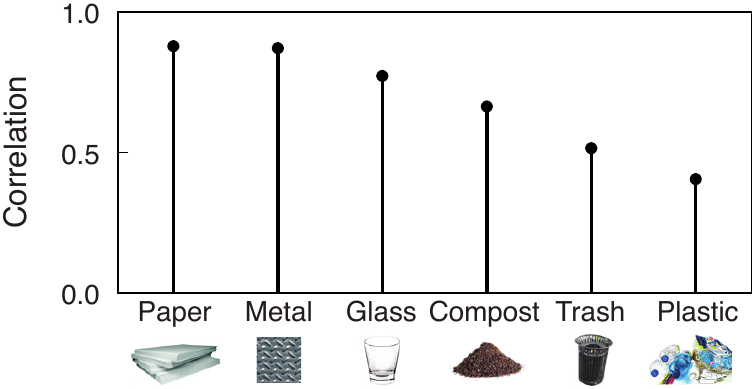}
	\figskip\caption{Correlations between human ratings and estimated ratings across all colors for each recycling-related concept using the Sector+Category model. The range of correlations is similar to fruits (Fig.~\ref{fig:E1Corr}.)}
	\figskip\label{fig:E3Corr}
\end{figure}

We tested the effectiveness of the Sector+Category model by correlating its estimates with mean human ratings over all 6 recycling-related concepts $\times$ 37 colors. The correlation was $r= .68$, $p < .001$, moderately strong, but significantly weaker than the corresponding correlation of .81 for fruit concepts in Experiment 1C ($z = 3.84$, $p< .001$). Fig.~\ref{fig:E3Corr} shows the correlations separately for each recycling concept. The fits range from .88 for paper to .40 for plastic, similar to the range for fruits in Experiment 1C (.94 to .49) (see Supplementary Fig.~\ref{fig:TestScatter}). 

\section{General discussion}
Creating color palettes that are semantically interpretable involves two main steps, (1) quantifying color-concept associations and (2) using those color-concept associations to generate unique assignments of colors to concepts for visualization. Our study focused on this first step, with the goal of understanding how to automatically estimate human color-concept associations from image statistics.

\subsection{Practical and theoretical applications}
We built on approaches from prior work \cite{lindner2012a, lindner2012b, lin2013, setlur2016, bartram2017} and harnessed perceptual and cognitive structure in color space to develop a new method for effectively estimating human color-concept associations. Our method can be used to create the input for various approaches to assigning colors to concepts for visualizations \cite{lin2013, setlur2016, havasi2010, bartram2017, schloss2018}. By  estimating full distributions of color-concept associations over color space that approximate human judgments (as opposed to identifying only the top associated colors), it should be possible to use assignment methods to define multiple candidate color palettes that are semantically interpretable for a given visualization. This flexibility will enable balancing semantics with other important factors in design, including perceptual discriminability \cite{healey1996, stone2014, szafir2018}, name difference \cite{heer2012}, aesthetics \cite{gramazio2017}, and emotional connotation \cite{bartram2017}. 

Our method can also be used to design stimuli for studies on visual reasoning. For example, evidence suggests people use assignment inference to interpret visualizations \cite{schloss2018} (Section \ref{sec:assigning_colors}), but little is known about how assignment inference works. Studying this processes requires the ability to carefully manipulate color-concept association strengths within visualizations, which requires having good estimates of human color-concept associations.

\subsection{Forming color-concept associations}
\begin{figure}[tb]
	\centering
	\includegraphics[width=0.8\columnwidth]{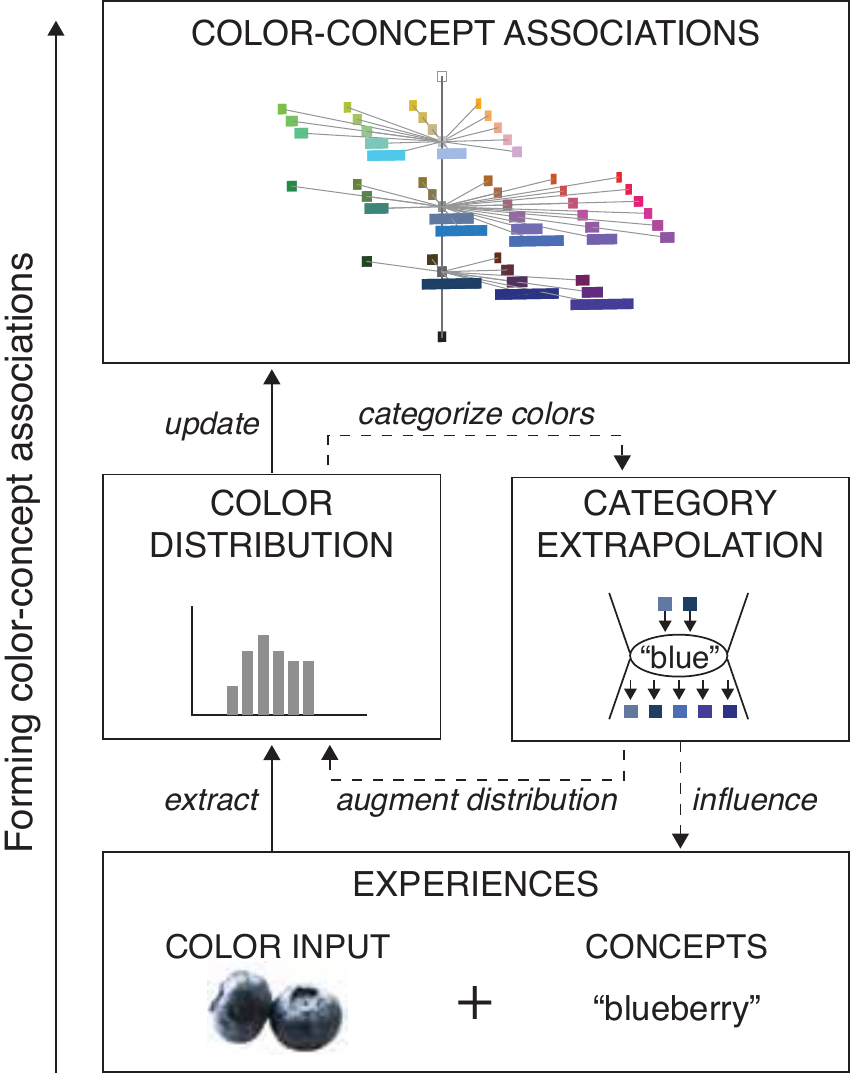}
	\figskip\caption{Process model for how color-concept associations are formed.}
	\figskip\label{fig:CogMod}
\end{figure}

In addition to providing a new method for estimating color-concept associations, this study sparked new insights into how people might form color-concept associations in the first place. In Fig.~\ref{fig:CogMod}, the path with solid arrows illustrates our initial premise for how people learn color concept associations from their environment. When they experience co-occurrences between colors and concepts, they extract color distributions and use them to update color-concept associations \cite{schlossPICS}. In Fig.~\ref{fig:CogMod}, color-concept association strength is indicated using marker width (e.g., \textit{blueberry} is highly associated with certain shades of blue). However, in the present study we found that it was insufficient to only extract colors (or nearby colors) from  images to estimate human color-concept associations (Experiments 1A and 1B). We needed to extrapolate to other colors that shared the same category as colors within the image to produce more accurate estimates (Experiment 1C). 

Based on these results, we propose there is another part of the process---category extrapolation---illustrated by the path with dashed arrows in Fig.~\ref{fig:CogMod}. While extracting the color distribution from color input, people categorize colors using basic color terms (e.g., ``blue''). This categorization process extrapolates to colors that are not in the visual input, but are within the same color category (e.g., extrapolating to all colors categorized as ``blue'' upon seeing a blueberry, even though only a subset of blues are found in the image). We believe that extrapolated colors augment the color distribution extracted from color input, which in turn further updates color-concept associations. Given that categorization can influence color perception \cite{kay1984,roberson2005,winawer2007,forder2019} and memory \cite{bae2015,kelly2017} (see \cite{witzel2018} for a review), category extrapolation may also feedback to influence color experiences.

\subsection{Open questions and limitations}

\parsec{Generalizability to other concepts.} In this study, we focused on fruit---concrete objects with directly observable colors---so we could study different methods of extracting and extrapolating colors from images where the colors would be systematic. We assessed generalizability to other, recycling-related concepts that are less color diagnostic \cite{tanaka1999} than fruit (e.g., paper, plastic, and glass can be any color), but recyclables are still concrete objects. However, there is concern that image-based methods may not be effective for estimating color-concept associations for abstract concepts that do not have directly observable colors \cite{lin2013, setlur2016}, though see \cite{bartram2017}. Nonetheless, people do have systematic associations between colors and abstract concepts \cite{osgood1957, wright1962, ou2004}. Future work will be needed to asses the boundary conditions of image based methods and further explore incorporating other, possibly language-based methods \cite{havasi2010, setlur2016} for estimating color-concept associations for abstract concepts.

\parsec{Image segmentation.} Our models might also be limited in their ability to generalize for concepts that refer to backgrounds rather than objects (e.g., ``sky'') \cite{lin2013}. All three models included a feature that extracted colors from figural regions and segmented away the backgrounds. Further research is needed to evaluate performance for background-related concepts, but limitations might be mitigated using semantic segmentation, in which particular regions of images are tagged with semantic labels\cite{semantic_segmentation}.

\parsec{Cultural differences.}
Our category extrapolation hypothesis implies that color-concept associations could differ between cultures whose languages have different color terms. Different languages partition color space in different ways \cite{berlin1969, gibson2017, roberson2005, kay2003, witzel2018}---e.g., some languages have separate color terms for blues and greens, whereas others have one term for both blues and greens. If a language has separate terms for blues and greens, experiencing blue objects like blueberries should result in color-concept associations that extrapolate only to other blues, not greens. But, if a language has one term for blues \textit{and} greens, experiencing blueberries should result in associations that extrapolate to blues \textit{and} greens. This is an exciting area for future research. 

\parsec{Structure of color categories.}
Our model defined color categories using a boundary approach---either a color was in a given category or not, with no distinction among category members. However, color categories have more complex structure, including a prototype, or best example, and varying levels of membership surrounding the prototype \cite{rosch1973}. A model that accounts for these complexities in category structure may improve on the fit to human color-concept associations.  

\section{Conclusion}
The goal of this study was to assess methods for automatically estimating color-concept associations from images. We tested different color extraction features that varied in color tolerance and spatial window, different kinds of images, and different concept sets. The most effective model used features that were relevant to human perception and cognition---features aligned with perceptual dimensions of color space and a feature that extrapolated to all colors within a color category. This model performed similarly well across the top 50 images from Google Images, curated photographs, and curated cartoon images. The model also generalized reasonably well to a different set of colors and concepts without changing any parameters. Through this study, we produced a method trained and validated on human data for automatically estimating color-concept associations, while generating new hypotheses about how color input and category extrapolation work together to produce human color-concept associations.

\acknowledgments{
The authors thank Christoph Witzel, Brian Yin, John Curtain, Joris Roos, Anna Bartel, and Emily Ward for their thoughtful feedback on this work, and Melissa Schoenlein, Shannon Sibrel, Autumn Wickman, Yuke Liang, and Marin Murack for their help with data collection. 
This work was supported in part by the Office of the Vice Chancellor for Research and Graduate Education at UW--Madison and the Wisconsin Alumni Research Foundation. The funding bodies played no role in designing the study, collecting, analyzing, or interpreting the data, or writing the manuscript.}

% \clearpage
% \tableofcontents

% bibliography on a separate page
\clearpage

\bibliographystyle{abbrv}

\bibliography{Vis2019_ColorAssoc}

%%%%%%%%%%%%%%%%%%%%%%%%%%%%%%%%%%%%%%%%%%%%%%%%%%%%%%%%%%%%%%%%%%%%%%%%%%%%
% supplementary material on separate page

%\begin{comment}
\clearpage
\appendix
\renewcommand*{\thesection}{S}
\counterwithin{figure}{section}
\counterwithin{table}{section}
\onecolumn
\section{Supplementary Material}\label{sec:supplementary}

Each section of the Supplementary Material is described below, and the Supplementary figures and tables follow after.

\subsection{Specifying the colors} \label{sec:defining_colors}
We defined the University of Wisconsin 58 (UW-58) colors using the following steps. First, we superimposed a 3D grid in CIELAB space with a distance $\Delta = 25$ in each dimension (L*, a*, and b*), and then selected all points whose colors could be rendered in RGB space (using MATLAB's \texttt{lab2rgb} function). We chose $\Delta = 25$ because it enabled us to achieve similar criteria that had been used to sample the Berkeley Color Project 37 colors (BCP-37), which are often used in color cognition studies \cite{palmer2010, schloss2013}. These criteria are: (a) include highly saturated approximations of unique hues (red, yellow, green, and blue) and intermediary hues, (b) include each hue sampled at three lightness levels and multiple chroma levels, and (c) include white, black, and intermediate grays at each lightness level. Unlike the BCP-37 colors, our color set is uniformly sampled in CIELAB space. Given the irregular shape of CIELAB space (and perceptual color spaces more generally), it is not possible to obtain colors for each hue at all possible lightness and chroma levels. We tried rotating the grid around the achromatic axis to get as many RGB valid high-chroma colors as possible and found a rotation of 3\degree\, included the largest set. Table \ref{table:UW_58_colors} contains the specific coordinates for all 58 UW colors in CIE 1931 xyY, CIELAB (L*,a*,b*) and CIELch (L*,c*,h) color spaces.\\
\\
Table \ref{table:BCP_37_colors} shows the coordinates for the BCP-37 colors in CIE 1931 xyY, CIELAB (L*,a*,b*) and CIELch (L*,c*,h) color spaces. The colors include eight hues (\textbf{R}ed, \textbf{O}range, \textbf{Y}ellow, c\textbf{H}artreuse, \textbf{G}reen, \textbf{C}yan, \textbf{B}lue, and \textbf{P}urple) at 4 saturation/lightness levels (\textbf{S}aturated, \textbf{L}ight, \textbf{M}uted, and \textbf{D}ark), plus Black (BK), dark gray (A1), medium gray (A2), light gray (A3) and white (WH). The color coordinates for SY, SG, SC and WH were slightly modified to fit into the RGB gamut assumed by the MATLAB function \texttt{lab2rgb()}. Chroma (c*) was reduced for SG from 64.77 to 59.7 ($\Delta E$ = 5.07), for SC from 44.73 to 43.2 ($\Delta E$ = 1.53), and for WH from 1.14 to 0 ($\Delta E$ = 1.14). Both lightness (L*) and chroma were reduced for SY with L* from 91.08 to 89 and c* from 86.87 to 85 ($\Delta E$ = 2.80).

\subsection{Results from human ratings of color-concept associations.}
\begin{figure}[ht]
	\centering
	\includegraphics[width=0.9\textwidth]{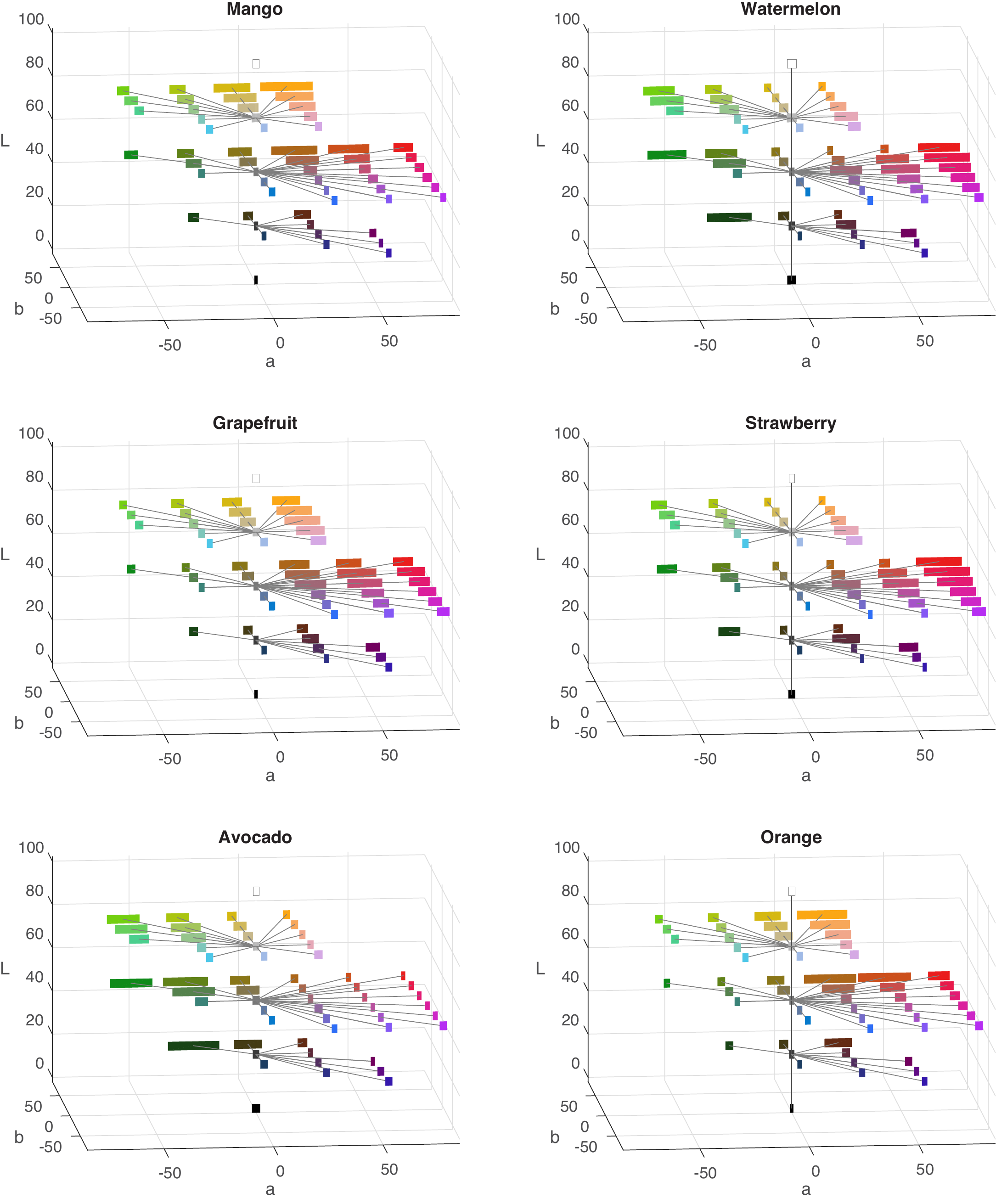}
	\figskip\caption{Mean color-concept associations for each of the UW-58 colors and each fruit, plotted in CIELAB space.  The width of the bars are proportional to association strength. Half of the fruit are shown here and the other half are in Supplementary Fig.~\ref{fig:FruitTree2}. The numerical values are in Table \ref{table:human_ratings} and in our github repository \url{https://github.com/Raginii/Color-Concept-Associations-using-Google-Images}.}
	\label{fig:FruitTree1}
\end{figure}

\begin{figure}[ht]
	\centering
	\includegraphics[width=0.9\textwidth]{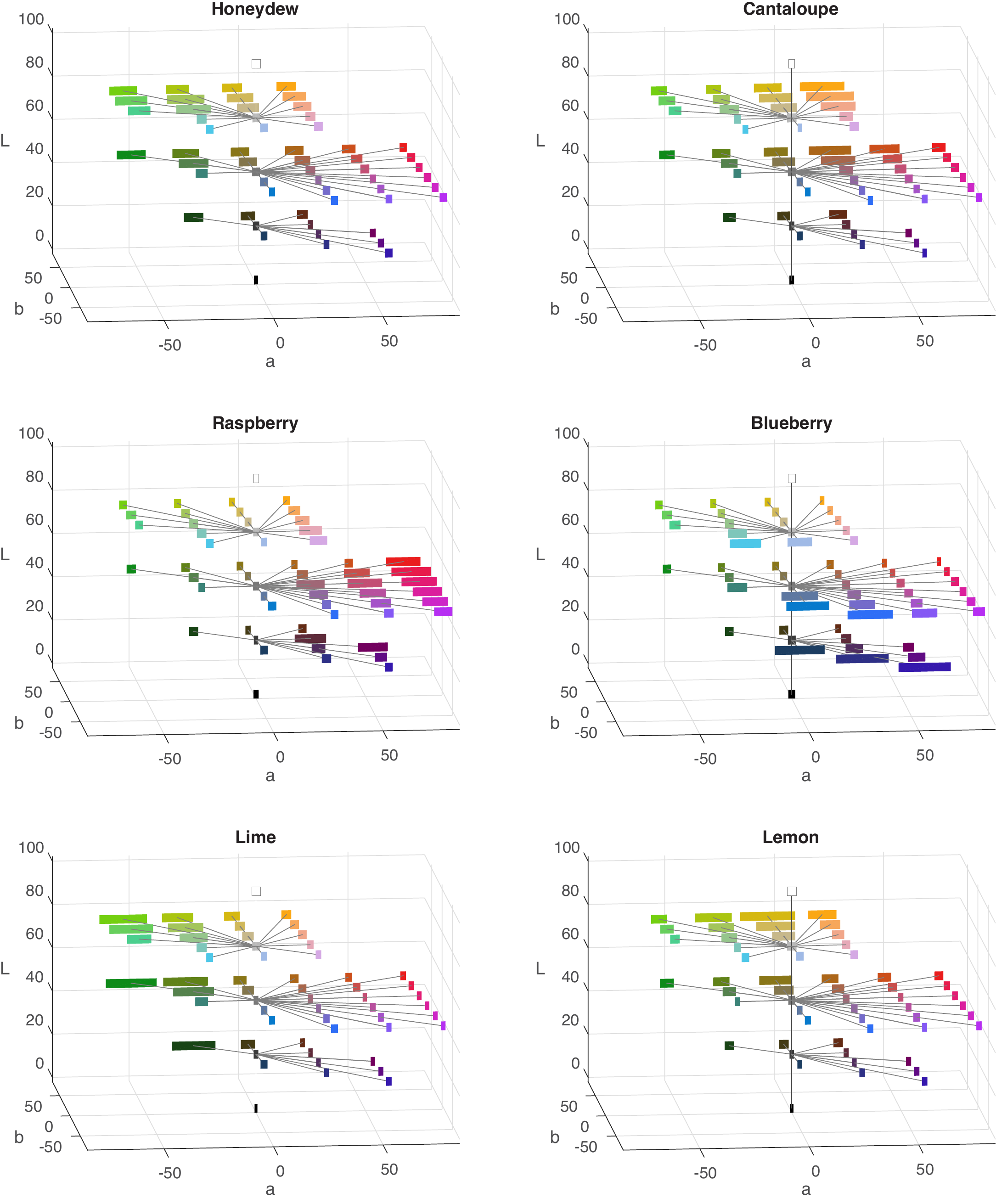}
	\figskip\caption{Mean color-concept associations for each of the UW-58 colors and each fruit, plotted in CIELAB space. The width of the bars are proportional to association strength. Half of the fruit are shown here and the other half are in Supplementary Fig.~\ref{fig:FruitTree1}. The numerical values are in Table \ref{table:human_ratings} and in our github repository \url{https://github.com/Raginii/Color-Concept-Associations-using-Google-Images}.}
	\label{fig:FruitTree2}
\end{figure}

Figures~\ref{fig:FruitTree1} and~\ref{fig:FruitTree2} show the mean human color-concept association ratings for the 12 different fruits and UW-58 colors. The mean ratings (range from 0 to 1) over all 54 participants can be found in Table \ref{table:human_ratings}. The full dataset including ratings from individual participants can be found at \url{https://github.com/Raginii/Color-Concept-Associations-using-Google-Images/blob/master/HumanRatingsData.csv}.

\subsection{Model fits for each fruit concepts in Experiment 1 and recycling concepts in Experiment 3}
\begin{figure*}[htb]
	\centering
	\includegraphics[width=1.0\textwidth]{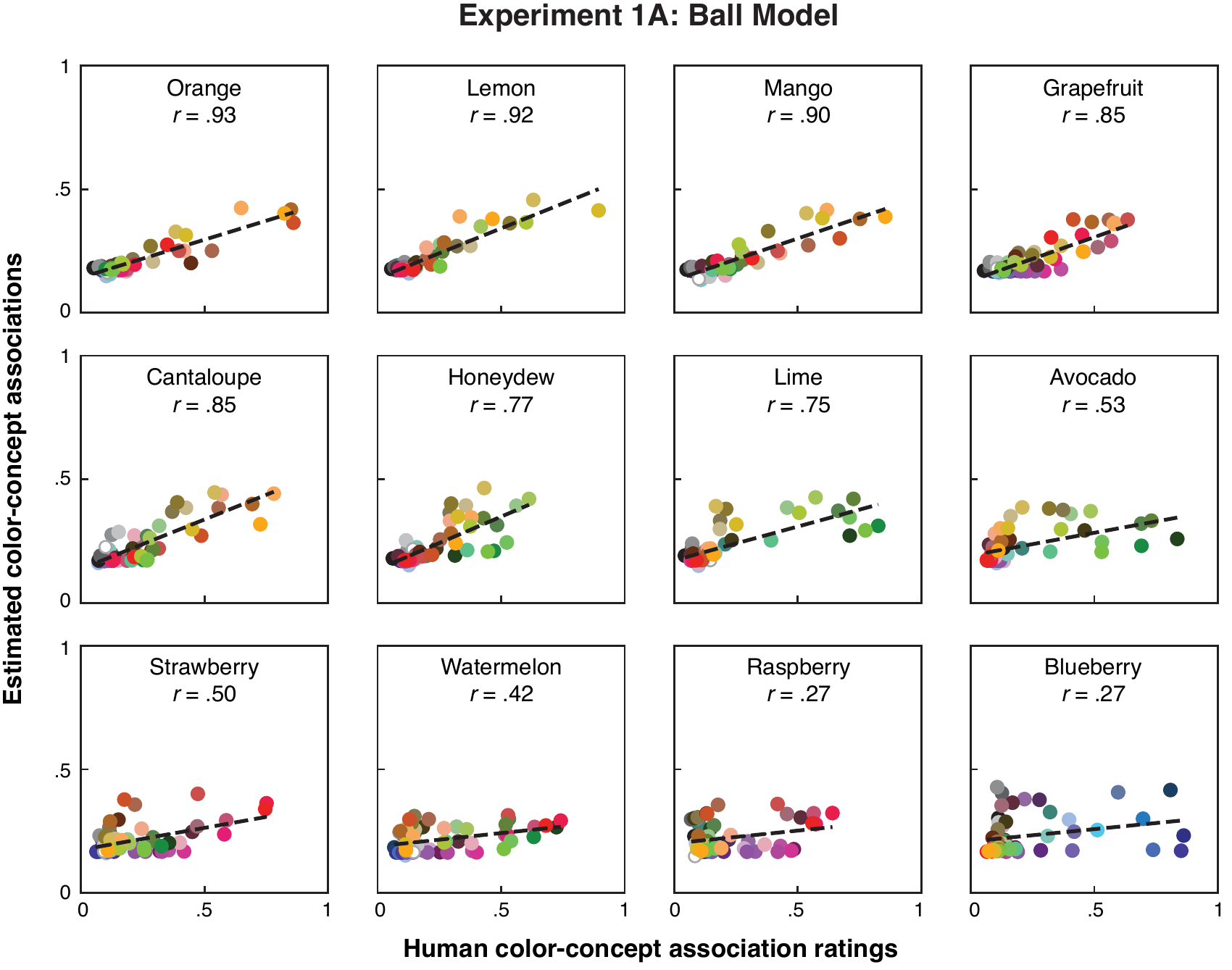}
	\figskip\caption{Correlations between mean human color-concept association ratings and estimates of the Ball model (Experiment 1A) across the UW-58 colors for each fruit.}
	\label{fig:AllScatter1}
\end{figure*}

\begin{figure*}[htb]
	\centering
	\includegraphics[width=1.0\textwidth]{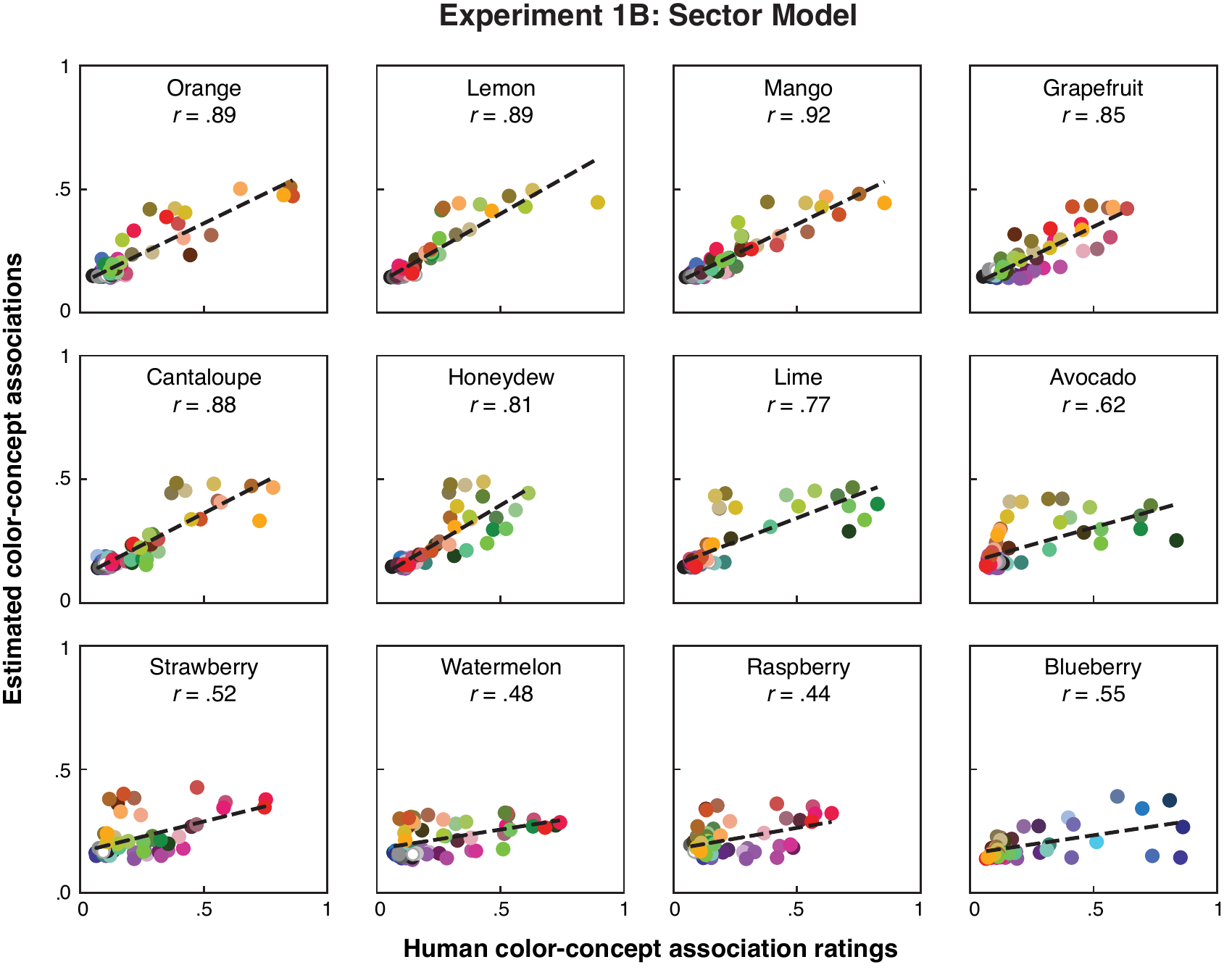}
	\figskip\caption{Correlations between mean human color-concept association ratings and estimates of the Sector model (Experiment 1B) across the UW-58 colors for each fruit.}
	\label{fig:AllScatter2}
\end{figure*}

\begin{figure*}[htb]
	\centering
	\includegraphics[width=1.0\textwidth]{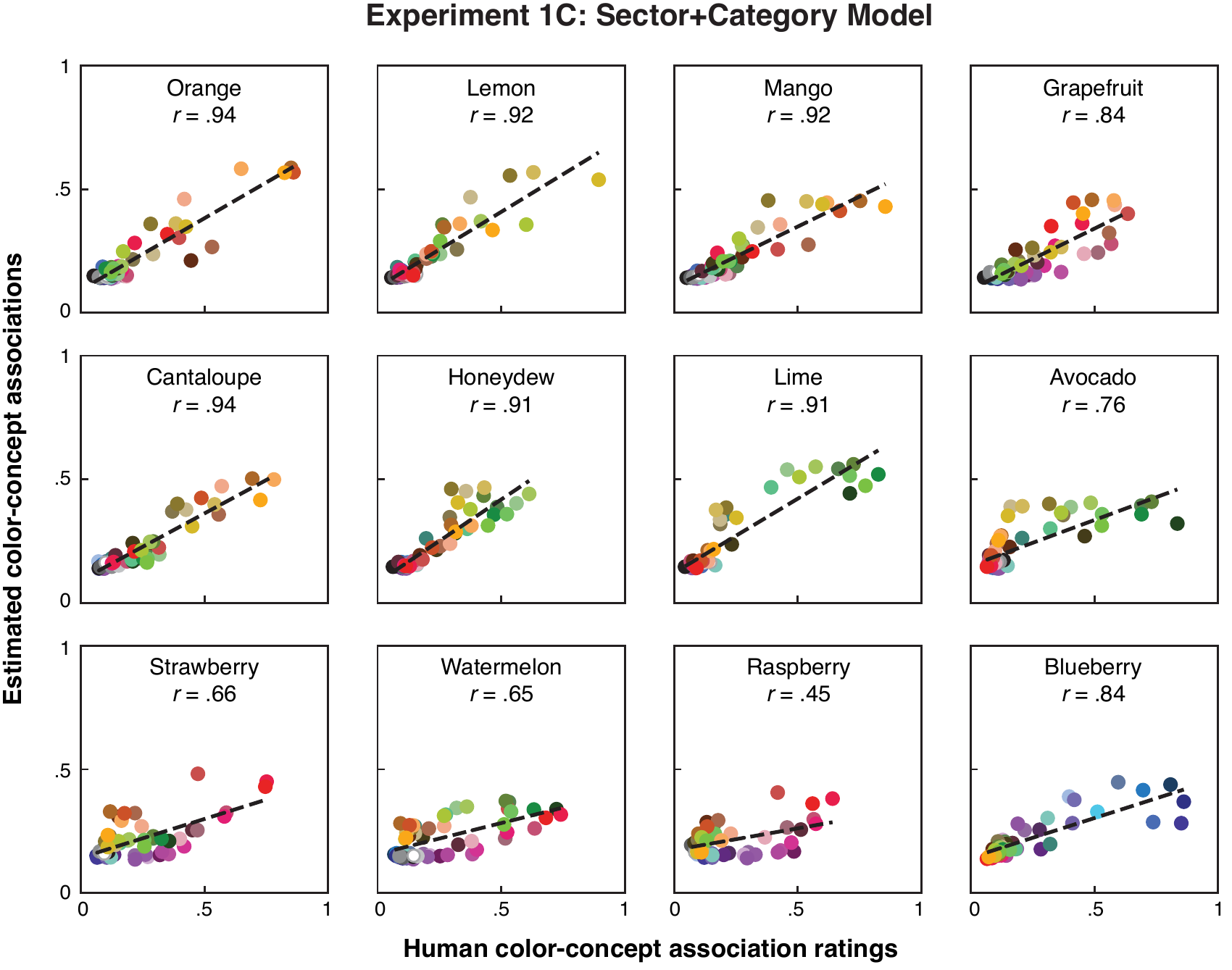}
	\figskip\caption{Correlations between mean human color-concept association ratings and estimates of the Sector+Category model (Experiment 1C) across the UW-58 colors for each fruit.}
	\label{fig:AllScatter3}
\end{figure*}

\begin{figure*}[ht]
\centering
	\includegraphics[width=0.7\textwidth]{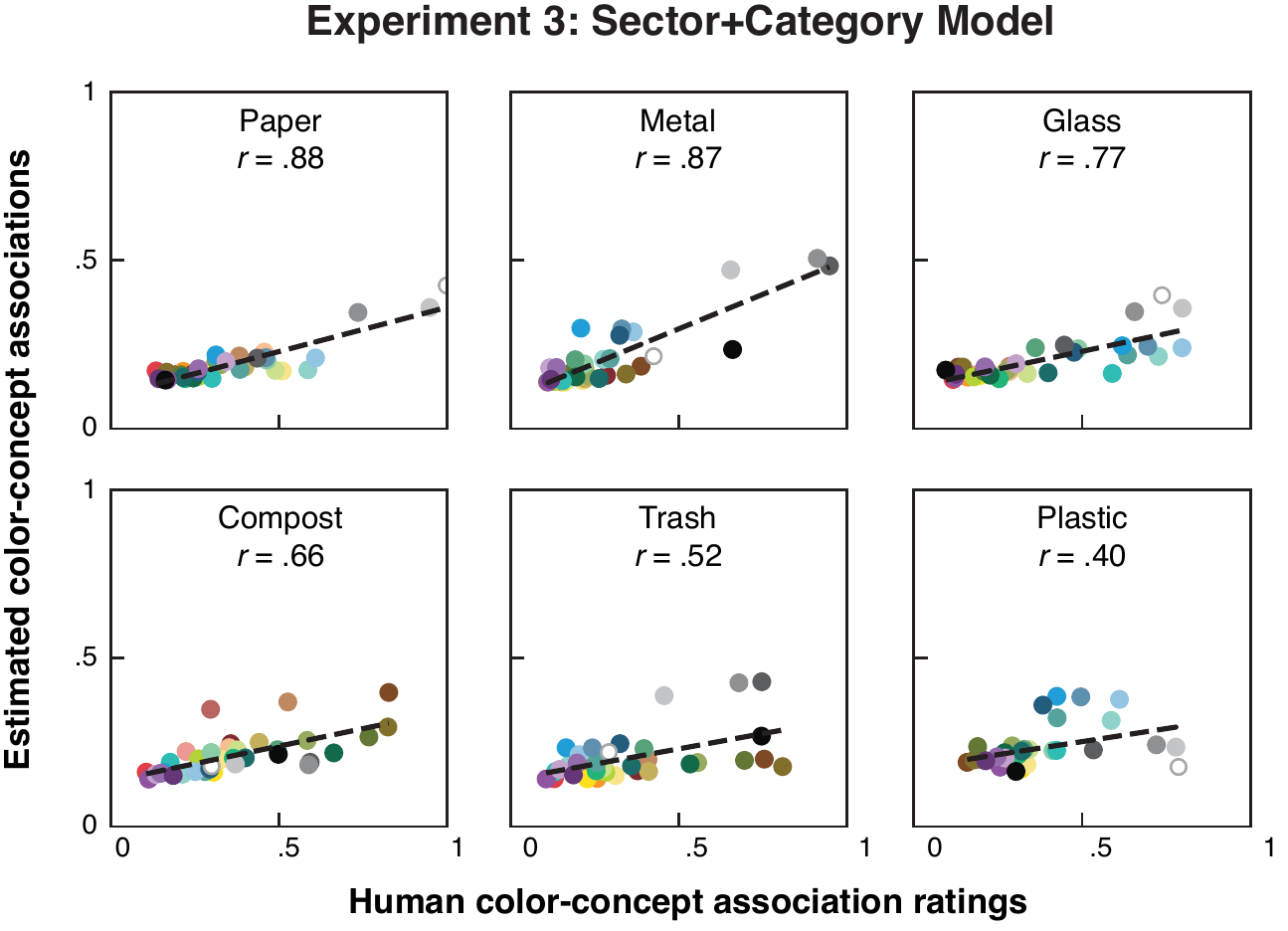}
	\figskip\caption{Correlations between mean human color-concept association ratings and estimates of the Sector+Category model (Experiment 3) across the BC-P37 colors for each recycling concept.}
	\label{fig:TestScatter}
\end{figure*}

Figures~\ref{fig:AllScatter1}--\ref{fig:AllScatter3} show the relation between model estimates (y-axis) and mean color-concept association ratings (x-axis) for each UW-58 color. These scatter plots and corresponding correlations are shown separately for the Ball model in Experiment 1A (Figure \ref{fig:AllScatter1}), Sector model in Experiment 1B (Figure \ref{fig:AllScatter2}) and Sector+Category in Experiment 1C (Figure \ref{fig:AllScatter3}).\\
\\
Figure~\ref{fig:TestScatter} shows the similar scatter plots from applying our Sector+Category model trained on fruit to a different color-concept association dataset for recycling concepts and the BCP-37 colors. 

\subsection{Varying the number of training images used} \label{sec:varying_number_of_images}

\begin{figure}[htb]
	\centering
	\includegraphics[width=0.65\columnwidth]{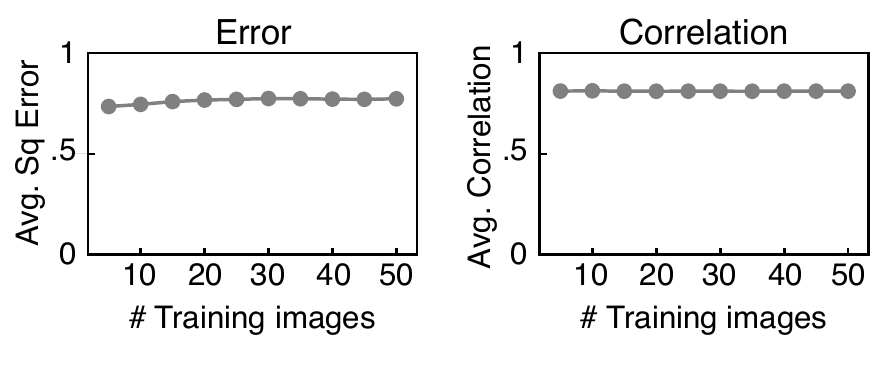}
	\figskip\caption{Variation in error and correlation with the number of training images used per fruit category. Using a larger set of training images from Google Images does not appreciably improve the performance of our model. The reason for this counterintuitive phenomenon may be that Google Images are automatically curated, so using as few as 5 images yields a highly representative sample.}
	\label{fig:errors}
\end{figure}

Figure~\ref{fig:errors} we shows the effect on mean squared test error of using a different number of training images when training the fruit model. Throughout our study, we used 50 Google Image search results, but as seen in the figure, it appears there is only a modest price to pay for using far fewer training images. While ``more data'' is often better, Google Image search results are already automatically curated; they are not random samples. This might explain why using as few as five search results produces a highly representative sample of the concept in question.

\begin{table*}[tb]
\sisetup{round-mode=places}
\centering
\figskip\caption{Coordinates for the University of Wisconsin 58 (UW-58) colors in CIE 1931 xyY space, CIELAB color space, and CIELch color space. The ``L*'' in CIELAB and CIELch is the same. The white point used to convert between CIE 1931 xyY and CIELAB space was CIE Illuminant D65 (x = 0.313, y = 0.329, Y = 100).}
\label{table:UW_58_colors}
\begin{tabular}
{|c*{1}{
    |S[round-precision=3]
    |S[round-precision=3]
    |S[round-precision=2]
    |S[round-precision=2]
    |S[round-precision=2]
    |S[round-precision=2]
    |S[round-precision=2]
    |S[round-precision=2]|
}}
\multicolumn{1}{c}{\textbf{Color}} &
\multicolumn{1}{c}{\textbf{x}} & \multicolumn{1}{c}{\textbf{y}} & \multicolumn{1}{c}{\textbf{Y}} & \multicolumn{1}{c}{\textbf{L*}} & \multicolumn{1}{c}{\textbf{a*}} & \multicolumn{1}{c}{\textbf{b*}} & \multicolumn{1}{c}{\textbf{c*}} & \multicolumn{1}{c}{\textbf{h}} \\ \hline
1 & 0.22397    & 0.28399    & 48.278     & 75         & -23.657    & -26.274    & 35.355     & 228.        \\ \hline
2 & 0.2081     & 0.21415    & 4.4155     & 25         & 1.3084     & -24.966    & 25.00         & 273.        \\ \hline
3 & 0.24471    & 0.25394    & 18.419     & 50         & 1.3084     & -24.966    & 25.00         & 273.        \\ \hline
4 & 0.26261    & 0.27354    & 48.278     & 75         & 1.3084     & -24.966    & 25.00         & 273.        \\ \hline
5 & 0.28644    & 0.19884    & 4.4155     & 25         & 26.274     & -23.657    & 35.355     & 318.        \\ \hline
6 & 0.29797    & 0.24051    & 18.419     & 50         & 26.274     & -23.657    & 35.355     & 318.        \\ \hline
7 & 0.30272    & 0.2622     & 48.278     & 75         & 26.274     & -23.657    & 35.355     & 318.        \\ \hline
8 & 0.36941    & 0.18108    & 4.4155     & 25         & 51.24      & -22.349    & 55.902     & 336.43     \\ \hline
9 & 0.35288    & 0.2259     & 18.419     & 50         & 51.24      & -22.349    & 55.902     & 336.43     \\ \hline
10 & 0.40795    & 0.21059    & 18.419     & 50         & 76.206     & -21.041    & 79.057     & 344.56     \\ \hline
11 & 0.18715    & 0.19157    & 18.419     & 50         & 2.6168     & -49.931    & 50.00      & 273.        \\ \hline
12 & 0.19063    & 0.1298     & 4.4155     & 25         & 27.583     & -48.623    & 55.902     & 299.57     \\ \hline
13 & 0.23146    & 0.18448    & 18.419     & 50         & 27.583     & -48.623    & 55.902     & 299.57     \\ \hline
14 & 0.25495    & 0.12284    & 4.4155     & 25         & 52.548     & -47.315    & 70.711     & 318.        \\ \hline
15 & 0.27871    & 0.17635    & 18.419     & 50         & 52.548     & -47.315    & 70.711     & 318.        \\ \hline
16 & 0.32783    & 0.1674     & 18.419     & 50         & 77.514     & -46.006    & 90.139     & 329.31     \\ \hline
17 & 0.17813    & 0.14021    & 18.419     & 50         & 28.891     & -73.589    & 79.057     & 291.43     \\ \hline
18 & 0.1742     & 0.082515   & 4.4155     & 25         & 53.857     & -72.28     & 90.139     & 306.69     \\ \hline
19 & 0.21726    & 0.13588    & 18.419     & 50         & 53.857     & -72.28     & 90.139     & 306.69     \\ \hline
20 & 0.25909    & 0.13088    & 18.419     & 50         & 78.822     & -70.972    & 106.07     & 318.        \\ \hline
21 & 0.25332    & 0.35108    & 18.419     & 50         & -24.966    & -1.3084    & 25.00      & 183.        \\ \hline
22 & 0.26938    & 0.34523    & 48.278     & 75         & -24.966    & -1.3084    & 25.00      & 183.        \\ \hline
23 & 0.31273    & 0.32902    & 0.00       & 0          & 0.00       & 0.00          & 0.00       & 0.00          \\ \hline
24 & 0.31273    & 0.32902    & 4.4155     & 25         & 0.00       & 0.00          & 0.00    & 0.00          \\ \hline
25 & 0.31273    & 0.32902    & 18.419     & 50         & 0.00       & 0.00          & 0.00    & 0.00          \\ \hline
26 & 0.31273    & 0.32902    & 48.278     & 75         & 0.00       & 0.00          & 0.00    & 0.00          \\ \hline
27 & 0.31273    & 0.32902    & 100.00     & 100        & 0.00       & 0.00          & 0.00          & 0.00          \\ \hline
28 & 0.41044    & 0.2905     & 4.4155     & 25         & 24.966     & 1.3084     & 25.        & 3.00          \\ \hline
29 & 0.37353    & 0.30534    & 18.419     & 50         & 24.966     & 1.3084     & 25.        & 3.00          \\ \hline
30 & 0.3568     & 0.31196    & 48.278     & 75         & 24.966     & 1.3084     & 25.        & 3.00          \\ \hline
31 & 0.43376    & 0.28095    & 18.419     & 50         & 49.931     & 2.6168     & 50.        & 3.00          \\ \hline
32 & 0.49181    & 0.25666    & 18.419     & 50         & 74.897     & 3.9252     & 75.        & 3.00          \\ \hline
33 & 0.3075     & 0.52435    & 4.4155     & 25         & -26.274    & 23.657     & 35.355     & 138.00        \\ \hline
34 & 0.31561    & 0.44408    & 18.419     & 50         & -26.274    & 23.657     & 35.355     & 138.00        \\ \hline
35 & 0.31654    & 0.41004    & 48.278     & 75         & -26.274    & 23.657     & 35.355     & 138.00        \\ \hline
36 & 0.27022    & 0.43268    & 48.278     & 75         & -51.24     & 22.349     & 55.902     & 156.43     \\ \hline
37 & 0.41792    & 0.44961    & 4.4155     & 25         & -1.3084    & 24.966     & 25.        & 93.00         \\ \hline
38 & 0.38179    & 0.40728    & 18.419     & 50         & -1.3084    & 24.966     & 25.        & 93.00         \\ \hline
39 & 0.36355    & 0.38634    & 48.278     & 75         & -1.3084    & 24.966     & 25.        & 93.00         \\ \hline
40 & 0.52174    & 0.37656    & 4.4155     & 25         & 23.657     & 26.274     & 35.355     & 48.00        \\ \hline
41 & 0.44682    & 0.36993    & 18.419     & 50         & 23.657     & 26.274     & 35.355     & 48.00         \\ \hline
42 & 0.41032    & 0.36214    & 48.278     & 75         & 23.657     & 26.274     & 35.355     & 48.00         \\ \hline
43 & 0.50874    & 0.33341    & 18.419     & 50         & 48.623     & 27.583     & 55.902     & 29.566     \\ \hline
44 & 0.56618    & 0.29873    & 18.419     & 50         & 73.589     & 28.891     & 79.057     & 21.435     \\ \hline
45 & 0.36753    & 0.52508    & 18.419     & 50         & -27.583    & 48.623     & 55.902     & 119.57     \\ \hline
46 & 0.35998    & 0.47135    & 48.278     & 75         & -27.583    & 48.623     & 55.902     & 119.57     \\ \hline
47 & 0.29736    & 0.57731    & 18.419     & 50         & -52.548    & 47.315     & 70.711     & 138.00        \\ \hline
48 & 0.31049    & 0.50294    & 48.278     & 75         & -52.548    & 47.315     & 70.711     & 138.00        \\ \hline
49 & 0.43671    & 0.47238    & 18.419     & 50         & -2.6168    & 49.931     & 50.         & 93.00         \\ \hline
50 & 0.4092     & 0.43925    & 48.278     & 75         & -2.6168    & 49.931     & 50.         & 93.00         \\ \hline
51 & 0.50246    & 0.42134    & 18.419     & 50         & 22.349     & 51.24      & 55.902     & 66.435     \\ \hline
52 & 0.4572     & 0.40735    & 48.278     & 75         & 22.349     & 51.24      & 55.902     & 66.435     \\ \hline
53 & 0.56315    & 0.37345    & 18.419     & 50         & 47.315     & 52.548     & 70.711     & 48.00        \\ \hline
54 & 0.61793    & 0.32963    & 18.419     & 50         & 72.28      & 53.857     & 90.139     & 36.69      \\ \hline
55 & 0.39381    & 0.52123    & 48.278     & 75         & -28.891    & 73.589     & 79.057     & 111.43     \\ \hline
56 & 0.34264    & 0.56147    & 48.278     & 75         & -53.857    & 72.28      & 90.139     & 126.69     \\ \hline
57 & 0.44388    & 0.48131    & 48.278     & 75         & -3.9252    & 74.897     & 75.         & 93.00         \\ \hline
58 & 0.49196    & 0.4425     & 48.278     & 75         & 21.041     & 76.206     & 79.057     & 74.565     \\ \hline
\end{tabular}
\end{table*}

\begin{table*}[tb]
\sisetup{round-mode=places}
\centering
\figskip \caption{Coordinates for the Berkeley Color Project 37 (BCP-37) colors in CIE 1931 xyY space, CIELAB color space, and CIELch color space. The ``L*'' in CIELAB and CIELch is the same. The white point used to convert between CIE 1931 xyY and CIELAB space was x = 0.312, y = 0.318, Y = 116. The colors include eight hues (\textbf{R}ed, \textbf{O}range, \textbf{Y}ellow, c\textbf{H}artreuse, \textbf{G}reen, \textbf{C}yan, \textbf{B}lue, and \textbf{P}urple) at 4 saturation/lightness levels (\textbf{S}aturated, \textbf{L}ight, \textbf{M}uted, and \textbf{D}ark), plus Black (BK), dark gray (A1), medium gray (A2), light gray (A3) and white (WH). The color coordinates for SY, SG, SC and WH were modified by decreasing the chroma c and lightness L values to fit the standard RGB gamut assumed by MATLAB.}

\label{table:BCP_37_colors}
\begin{tabular}{
|c*{1}{
    |S[round-precision=3]
    |S[round-precision=3]
    |S[round-precision=2]
    |S[round-precision=2]
    |S[round-precision=2]
    |S[round-precision=2]
    |S[round-precision=2]
    |S[round-precision=2]|
}}

%\textbf{x} & \textbf{y} & \textbf{Y} & \textbf{L} & \textbf{a} & \textbf{b} & \textbf{c} & \textbf{h} 
\multicolumn{1}{c}{\textbf{BCP Notation}} &
\multicolumn{1}{c}{\textbf{x}} & \multicolumn{1}{c}{\textbf{y}} & \multicolumn{1}{c}{\textbf{Y}} & \multicolumn{1}{c}{\textbf{L*}} & \multicolumn{1}{c}{\textbf{a*}} & \multicolumn{1}{c}{\textbf{b*}} & \multicolumn{1}{c}{\textbf{c*}} & \multicolumn{1}{c}{\textbf{h}}\\ \hline

SR           & 0.549 & 0.313 & 22.93 & 51.573 & 62.234     & 32.198    & 70.07    & 27.356 \\ \hline
LR           & 0.407 & 0.326 & 49.95 & 71.596 & 31.578     & 16.68     & 35.713   & 27.844 \\ \hline
MR           & 0.441 & 0.324 & 22.93 & 51.573 & 33.58      & 16.981    & 37.63    & 26.825 \\ \hline
DR           & 0.506 & 0.311 & 7.6   & 30.764 & 37.017     & 16.39     & 40.483   & 23.882 \\ \hline
SO           & 0.513 & 0.412 & 49.95 & 71.596 & 31.215     & 69.647    & 76.322   & 65.859 \\ \hline
LO           & 0.399 & 0.366 & 68.56 & 81.348 & 15.001     & 30.172    & 33.696   & 63.565 \\ \hline
MO           & 0.423 & 0.375 & 34.86 & 61.699 & 15.94      & 30.33     & 34.263   & 62.275 \\ \hline
DO           & 0.481 & 0.388 & 10.76 & 36.51  & 18.354     & 30.597    & 35.679   & 59.042 \\ \hline
SY           & 0.446 & 0.472 & 91.25 & 91.082 & -5.7506    & 86.678    & 86.868   & 93.796 \\ \hline
LY           & 0.391 & 0.413 & 91.25 & 91.082 & -5.4588    & 47.705    & 48.016   & 96.528 \\ \hline
MY           & 0.407 & 0.426 & 49.95 & 71.596 & -3.3302    & 45.936    & 46.057   & 94.147 \\ \hline
DY           & 0.437 & 0.45  & 18.43 & 46.827 & -0.92513   & 43.347    & 43.357   & 91.223 \\ \hline
SH           & 0.387 & 0.504 & 68.56 & 81.348 & -32.919    & 72.055    & 79.218   & 114.55 \\ \hline
LH           & 0.357 & 0.42  & 79.9  & 86.444 & -20.62     & 40.644    & 45.576   & 116.9  \\ \hline
MH           & 0.36  & 0.436 & 42.4  & 66.939 & -19.975    & 37.45     & 42.444   & 118.07 \\ \hline
DH           & 0.369 & 0.473 & 18.43 & 46.827 & -19.923    & 36.863    & 41.903   & 118.39 \\ \hline
SG           & 0.254 & 0.449 & 42.4  & 66.939 & -59.948    & 24.537    & 64.775   & 157.74 \\ \hline
LG           & 0.288 & 0.381 & 63.9  & 79.091 & -34.126    & 15.212    & 37.363   & 155.97 \\ \hline
MG           & 0.281 & 0.392 & 34.86 & 61.699 & -33.267    & 14.065    & 36.118   & 157.08 \\ \hline
DG           & 0.261 & 0.419 & 12.34 & 38.964 & -33.292    & 12.408    & 35.529   & 159.56 \\ \hline
SC           & 0.226 & 0.335 & 49.95 & 71.596 & -44.315    & -6.1068   & 44.734   & 187.85 \\ \hline
LC           & 0.267 & 0.33  & 68.56 & 81.348 & -26.118    & -2.7294   & 26.26    & 185.97 \\ \hline
MC           & 0.254 & 0.328 & 34.86 & 61.699 & -25.402    & -4.1266   & 25.735   & 189.23 \\ \hline
DC           & 0.233 & 0.324 & 13.92 & 41.216 & -24.26     & -5.4518   & 24.865   & 192.67 \\ \hline
SB           & 0.2   & 0.23  & 34.86 & 61.699 & -13.209    & -38.399   & 40.608   & 251.02 \\ \hline
LB           & 0.255 & 0.278 & 59.25 & 76.726 & -8.8676    & -20.82    & 22.63    & 246.93 \\ \hline
MB           & 0.241 & 0.265 & 28.9  & 56.991 & -7.8584    & -21.411   & 22.807   & 249.85 \\ \hline
DB           & 0.212 & 0.236 & 10.76 & 36.51  & -6.5572    & -23.727   & 24.616   & 254.55 \\ \hline
SP           & 0.272 & 0.156 & 18.43 & 46.827 & 57.212     & -50.49    & 76.305   & 318.57 \\ \hline
LP           & 0.29  & 0.242 & 49.95 & 71.596 & 26.028     & -27.872   & 38.135   & 313.04 \\ \hline
MP           & 0.287 & 0.222 & 22.93 & 51.573 & 28.052     & -27.816   & 39.505   & 315.24 \\ \hline
DP           & 0.28  & 0.181 & 7.6   & 30.764 & 33.038     & -29.66    & 44.399   & 318.08 \\ \hline
BK           & 0.31  & 0.316 & 0.3   & 2.3361 & -0.0012256 & -0.069311 & 0.069322 & 268.99 \\ \hline
A1           & 0.31  & 0.316 & 12.34 & 38.964 & -0.0096122 & -0.5405   & 0.54058  & 268.98 \\ \hline
A2           & 0.31  & 0.316 & 31.88 & 59.419 & -0.013189  & -0.74164  & 0.74176  & 268.98 \\ \hline
A3           & 0.31  & 0.316 & 63.9  & 79.091 & -0.01663   & -0.93509  & 0.93524  & 268.98 \\ \hline
WH           & 0.31  & 0.316 & 116.   & 100.00    & -0.020286  & -1.1407   & 1.1409   & 268.98 \\ \hline
\end{tabular}
\end{table*}

\begin{table*}[tb]
\sisetup{round-mode=places}
\centering
\figskip \caption{Average color ratings obtained from humans for all UW-58 colors and fruit concepts. The fruit concepts shown in the table are Mango, Watermelon (Waterm.), Honeydew (Honeyd.), Cantaloupe (Cantal.), Grapefruit (Grapefr.), Strawberry (Strawb.), Raspberry (Raspb.), Blueberry (Blueb.), Avocado, Orange, Lime and Lemon.}
\label{table:human_ratings}
\begin{tabular}%{|c|c|c|c|}
{|c*{1}{
    |S[round-precision=2]
    |S[round-precision=2]
    |S[round-precision=2]
    |S[round-precision=2]
    |S[round-precision=2]
    |S[round-precision=2]
    |S[round-precision=2]
    |S[round-precision=2]
    |S[round-precision=2]
    |S[round-precision=2]
    |S[round-precision=2]
    |S[round-precision=2]|
}}
\multicolumn{1}{c}{\textbf{Color}} &
\multicolumn{1}{c}{\textbf{Mango}} & \multicolumn{1}{c}{\textbf{Waterm.}} & \multicolumn{1}{c}{\textbf{Honeyd.}} & \multicolumn{1}{c}{\textbf{Cantal.}} & \multicolumn{1}{c}{\textbf{Grapefr.}} & \multicolumn{1}{c}{\textbf{Strawb.}} & \multicolumn{1}{c}{\textbf{Raspb.}} & \multicolumn{1}{c}{\textbf{Blueb.}} & \multicolumn{1}{c}{\textbf{Avocado}} &
\multicolumn{1}{c}{\textbf{Orange}} &
\multicolumn{1}{c}{\textbf{Lime}} &
\multicolumn{1}{c}{\textbf{Lemon}} \\ \hline
% & mango & watermelon & honeydew & cantaloupe & grapefruit & strawberry & raspberry & blueberry & avocado & orange & lime & lemon \\
1 & 0.111435185 & 0.092314815 & 0.121666667 & 0.092037037 & 0.091759259 & 0.100787037 & 0.126527778 & 0.512222222 & 0.111944444 & 0.09712963 & 0.108009259 & 0.12087963 \\ \hline
2 & 0.076203704 & 0.067962963 & 0.106203704 & 0.094490741 & 0.083564815 & 0.082268519 & 0.117685185 & 0.808055556 & 0.116805556 & 0.07837963 & 0.109907407 & 0.075185185 \\ \hline
3 & 0.107685185 & 0.129814815 & 0.128981481 & 0.101944444 & 0.127222222 & 0.094907407 & 0.106851852 & 0.596435185 & 0.101296296 & 0.080555556 & 0.106064815 & 0.084027778 \\ \hline
4 & 0.108287037 & 0.11287037 & 0.127037037 & 0.070555556 & 0.115555556 & 0.1025 & 0.093657407 & 0.397916667 & 0.104398148 & 0.102731481 & 0.098981481 & 0.127175926 \\ \hline
5 & 0.102314815 & 0.101527778 & 0.084768519 & 0.094537037 & 0.186111111 & 0.100833333 & 0.220833333 & 0.278240741 & 0.100185185 & 0.092083333 & 0.074027778 & 0.076296296 \\ \hline
6 & 0.115277778 & 0.185231481 & 0.098842593 & 0.133472222 & 0.230462963 & 0.218101852 & 0.3 & 0.2175 & 0.130925926 & 0.100462963 & 0.0775 & 0.107175926 \\ \hline
7 & 0.108888889 & 0.212083333 & 0.136435185 & 0.151666667 & 0.253333333 & 0.269814815 & 0.282916667 & 0.129027778 & 0.13337963 & 0.14 & 0.089444444 & 0.120324074 \\ \hline
8 & 0.111435185 & 0.252685185 & 0.093425926 & 0.102592593 & 0.230046296 & 0.318703704 & 0.484259259 & 0.179722222 & 0.066574074 & 0.114259259 & 0.069259259 & 0.080694444 \\ \hline
9 & 0.16 & 0.38337963 & 0.098611111 & 0.129074074 & 0.36537037 & 0.355138889 & 0.430231481 & 0.136018519 & 0.07162037 & 0.121851852 & 0.090833333 & 0.089537037 \\ \hline
10 & 0.148611111 & 0.400972222 & 0.112083333 & 0.094305556 & 0.297268519 & 0.417222222 & 0.473888889 & 0.108564815 & 0.078703704 & 0.151342593 & 0.067546296 & 0.090648148 \\ \hline
11 & 0.103564815 & 0.076296296 & 0.094953704 & 0.086666667 & 0.098888889 & 0.069583333 & 0.136805556 & 0.69712963 & 0.09212963 & 0.093101852 & 0.093888889 & 0.097083333 \\ \hline
12 & 0.095555556 & 0.075648148 & 0.093287037 & 0.078009259 & 0.089166667 & 0.062962963 & 0.152916667 & 0.861666667 & 0.113564815 & 0.09587963 & 0.071481481 & 0.090833333 \\ \hline
13 & 0.080277778 & 0.089537037 & 0.120277778 & 0.096851852 & 0.13625 & 0.10587963 & 0.149444444 & 0.418611111 & 0.119259259 & 0.090046296 & 0.100787037 & 0.09037037 \\  \hline
14 & 0.068240741 & 0.088842593 & 0.100185185 & 0.081203704 & 0.161527778 & 0.133611111 & 0.20037037 & 0.283472222 & 0.109907407 & 0.083472222 & 0.076990741 & 0.09125 \\ \hline
15 & 0.144768519 & 0.201944444 & 0.114722222 & 0.097175926 & 0.25912037 & 0.260740741 & 0.328518519 & 0.185833333 & 0.089212963 & 0.093148148 & 0.091666667 & 0.104907407 \\ \hline
16 & 0.124814815 & 0.283148148 & 0.103333333 & 0.107685185 & 0.223888889 & 0.327638889 & 0.419537037 & 0.143101852 & 0.078287037 & 0.133194444 & 0.073101852 & 0.092361111 \\ \hline
17 & 0.092777778 & 0.088101852 & 0.109537037 & 0.081018519 & 0.107314815 & 0.074074074 & 0.126018519 & 0.737824074 & 0.088842593 & 0.087546296 & 0.110324074 & 0.095972222 \\ \hline
18 & 0.082361111 & 0.101342593 & 0.115694444 & 0.0725 & 0.108009259 & 0.065092593 & 0.12162037 & 0.851527778 & 0.11162037 & 0.086574074 & 0.086018519 & 0.087453704 \\ \hline
19 & 0.093564815 & 0.118009259 & 0.107222222 & 0.092962963 & 0.155231481 & 0.104398148 & 0.153287037 & 0.411157407 & 0.104675926 & 0.094027778 & 0.080648148 & 0.079351852 \\ \hline
20 & 0.093194444 & 0.144398148 & 0.116574074 & 0.092453704 & 0.2025 & 0.218518519 & 0.294814815 & 0.190046296 & 0.115648148 & 0.122824074 & 0.072037037 & 0.084027778 \\ \hline
21 & 0.114398148 & 0.150416667 & 0.19625 & 0.123564815 & 0.095046296 & 0.09375 & 0.100972222 & 0.320925926 & 0.207546296 & 0.089166667 & 0.206157407 & 0.079351852 \\ \hline
22 & 0.110092593 & 0.116990741 & 0.158518519 & 0.112824074 & 0.103703704 & 0.12037037 & 0.153611111 & 0.310925926 & 0.148564815 & 0.117685185 & 0.165648148 & 0.11125 \\ \hline
23 & 0.050601852 & 0.143240741 & 0.061944444 & 0.072314815 & 0.053888889 & 0.110138889 & 0.082037037 & 0.111851852 & 0.132268519 & 0.051481481 & 0.043564815 & 0.055694444 \\ \hline
24 & 0.066296296 & 0.101018519 & 0.101435185 & 0.078888889 & 0.078657407 & 0.08287037 & 0.068842593 & 0.126157407 & 0.102592593 & 0.083009259 & 0.072731481 & 0.068657407 \\ \hline
25 & 0.076805556 & 0.089166667 & 0.122777778 & 0.12462963 & 0.077222222 & 0.075601852 & 0.093796296 & 0.107083333 & 0.116759259 & 0.072407407 & 0.073101852 & 0.108842593 \\ \hline
26 & 0.143333333 & 0.160601852 & 0.115231481 & 0.152453704 & 0.107685185 & 0.094259259 & 0.088564815 & 0.11337963 & 0.09587963 & 0.102916667 & 0.118564815 & 0.152361111 \\ \hline
27 & 0.100138889 & 0.146064815 & 0.143240741 & 0.099537037 & 0.105648148 & 0.093472222 & 0.084259259 & 0.121990741 & 0.105092593 & 0.106111111 & 0.146388889 & 0.157037037 \\ \hline
28 & 0.112175926 & 0.326527778 & 0.083935185 & 0.139305556 & 0.269074074 & 0.450231481 & 0.512268519 & 0.168888889 & 0.074212963 & 0.122916667 & 0.067268519 & 0.080092593 \\ \hline
29 & 0.222592593 & 0.51625 & 0.155787037 & 0.225324074 & 0.514490741 & 0.470694444 & 0.456712963 & 0.125092593 & 0.089768519 & 0.176666667 & 0.085648148 & 0.106759259 \\ \hline
30 & 0.207361111 & 0.380416667 & 0.159583333 & 0.216388889 & 0.457962963 & 0.399444444 & 0.366574074 & 0.113194444 & 0.097824074 & 0.183842593 & 0.098657407 & 0.130324074 \\ \hline
31 & 0.185972222 & 0.635277778 & 0.145 & 0.186342593 & 0.568009259 & 0.587824074 & 0.564537037 & 0.125740741 & 0.075740741 & 0.184814815 & 0.067175926 & 0.08912037 \\ \hline
32 & 0.151898148 & 0.523935185 & 0.108101852 & 0.125462963 & 0.340972222 & 0.579953704 & 0.573194444 & 0.12412037 & 0.071527778 & 0.165092593 & 0.074675926 & 0.089212963 \\ \hline
33 & 0.177824074 & 0.72375 & 0.313425926 & 0.209398148 & 0.138055556 & 0.356203704 & 0.147407407 & 0.137268519 & 0.835555556 & 0.126990741 & 0.710648148 & 0.150925926 \\ \hline
34 & 0.254305556 & 0.531435185 & 0.48337963 & 0.267916667 & 0.138611111 & 0.291157407 & 0.161435185 & 0.13712963 & 0.689953704 & 0.129490741 & 0.664212963 & 0.218240741 \\ \hline
35 & 0.166157407 & 0.319675926 & 0.561712963 & 0.316481481 & 0.150925926 & 0.188611111 & 0.137453704 & 0.139490741 & 0.404351852 & 0.148333333 & 0.457268519 & 0.25287037 \\ \hline
36 & 0.153888889 & 0.275833333 & 0.362638889 & 0.205416667 & 0.133935185 & 0.156944444 & 0.129351852 & 0.181481481 & 0.32162037 & 0.119027778 & 0.39212963 & 0.151944444 \\ \hline
37 & 0.157037037 & 0.18287037 & 0.238009259 & 0.209212963 & 0.148333333 & 0.115787037 & 0.082222222 & 0.142314815 & 0.460740741 & 0.143796296 & 0.232777778 & 0.157222222 \\ \hline
38 & 0.276851852 & 0.149027778 & 0.291990741 & 0.369351852 & 0.192638889 & 0.114907407 & 0.095787037 & 0.109675926 & 0.373333333 & 0.209490741 & 0.186018519 & 0.320833333 \\ \hline
39 & 0.338009259 & 0.150833333 & 0.357453704 & 0.424861111 & 0.256990741 & 0.138564815 & 0.108935185 & 0.123009259 & 0.161296296 & 0.291111111 & 0.185231481 & 0.375648148 \\ \hline
40 & 0.274537037 & 0.126527778 & 0.17087963 & 0.285972222 & 0.181018519 & 0.151759259 & 0.13162037 & 0.088657407 & 0.156944444 & 0.444953704 & 0.076805556 & 0.152685185 \\ \hline
41 & 0.542824074 & 0.20712963 & 0.253425926 & 0.557037037 & 0.559537037 & 0.218101852 & 0.177592593 & 0.107268519 & 0.105648148 & 0.53 & 0.128611111 & 0.197407407 \\ \hline
42 & 0.428472222 & 0.269305556 & 0.293148148 & 0.569212963 & 0.582314815 & 0.245462963 & 0.228796296 & 0.102962963 & 0.097592593 & 0.417731481 & 0.136435185 & 0.197824074 \\ \hline
43 & 0.418703704 & 0.528009259 & 0.181342593 & 0.314768519 & 0.634953704 & 0.472824074 & 0.418564815 & 0.087546296 & 0.089027778 & 0.396574074 & 0.114166667 & 0.143796296 \\ \hline
44 & 0.174351852 & 0.741111111 & 0.12962963 & 0.132037037 & 0.450416667 & 0.750833333 & 0.640509259 & 0.085555556 & 0.081898148 & 0.217083333 & 0.08337963 & 0.102731481 \\ \hline
45 & 0.277268519 & 0.51875 & 0.428888889 & 0.292175926 & 0.124907407 & 0.292222222 & 0.13375 & 0.111296296 & 0.73125 & 0.122361111 & 0.72537037 & 0.26212963 \\ \hline
46 & 0.274259259 & 0.361111111 & 0.612962963 & 0.279027778 & 0.177407407 & 0.194861111 & 0.13912037 & 0.118935185 & 0.485092593 & 0.16337963 & 0.572175926 & 0.417962963 \\ \hline
47 & 0.226666667 & 0.632175926 & 0.470972222 & 0.252314815 & 0.131759259 & 0.325509259 & 0.149027778 & 0.129953704 & 0.690694444 & 0.100185185 & 0.825833333 & 0.223333333 \\ \hline
48 & 0.221435185 & 0.539861111 & 0.523009259 & 0.268935185 & 0.134166667 & 0.255277778 & 0.165 & 0.157731481 & 0.531064815 & 0.131157407 & 0.710138889 & 0.218842593 \\ \hline
49 & 0.38037037 & 0.137824074 & 0.297962963 & 0.389722222 & 0.248518519 & 0.097824074 & 0.104212963 & 0.10712963 & 0.317453704 & 0.281759259 & 0.210185185 & 0.535462963 \\ \hline
50 & 0.535231481 & 0.116527778 & 0.431666667 & 0.540648148 & 0.365694444 & 0.11962963 & 0.112546296 & 0.111805556 & 0.208564815 & 0.383564815 & 0.16912037 & 0.630138889 \\ \hline
51 & 0.751805556 & 0.092407407 & 0.297083333 & 0.692824074 & 0.489861111 & 0.118009259 & 0.09787037 & 0.113055556 & 0.123287037 & 0.849722222 & 0.13537037 & 0.269166667 \\ \hline
52 & 0.618101852 & 0.14337963 & 0.379398148 & 0.780324074 & 0.578611111 & 0.163657407 & 0.190787037 & 0.086944444 & 0.119675926 & 0.648009259 & 0.147731481 & 0.332685185 \\ \hline
53 & 0.670231481 & 0.128935185 & 0.220462963 & 0.486574074 & 0.414675926 & 0.175694444 & 0.133333333 & 0.075 & 0.081851852 & 0.859490741 & 0.107175926 & 0.217268519 \\ \hline
54 & 0.315833333 & 0.680972222 & 0.113888889 & 0.215694444 & 0.325277778 & 0.745694444 & 0.559907407 & 0.065324074 & 0.065416667 & 0.348981481 & 0.090092593 & 0.143657407 \\ \hline
55 & 0.262083333 & 0.268935185 & 0.375277778 & 0.24412037 & 0.204398148 & 0.151527778 & 0.111018519 & 0.097916667 & 0.365555556 & 0.170555556 & 0.506342593 & 0.601435185 \\ \hline
56 & 0.197175926 & 0.511296296 & 0.446574074 & 0.266759259 & 0.125277778 & 0.25625 & 0.130694444 & 0.114398148 & 0.529814815 & 0.12125 & 0.773842593 & 0.253101852 \\ \hline
57 & 0.599814815 & 0.112268519 & 0.324814815 & 0.449444444 & 0.323194444 & 0.108287037 & 0.093518519 & 0.096759259 & 0.15037037 & 0.424444444 & 0.251527778 & 0.894444444 \\ \hline
58 & 0.854212963 & 0.115138889 & 0.31587963 & 0.725740741 & 0.454166667 & 0.108564815 & 0.110138889 & 0.072638889 & 0.11162037 & 0.823194444 & 0.15962963 & 0.464953704 \\ \hline
\end{tabular}
\end{table*}
%\end{comment}
\end{document}